\documentclass{aa}  
\usepackage[utf8]{inputenc}
\usepackage{slashbox}
\usepackage{natbib}
\usepackage[colorlinks]{hyperref}
\usepackage{xcolor}
\usepackage{ulem}
\usepackage{booktabs}
\usepackage{array}
\usepackage{adjustbox}
\usepackage{arydshln}
\hypersetup{ citecolor=[RGB]{44, 117, 255}}
\usepackage{amsmath}
\usepackage{listings,comment}

\def\rosat{ROSAT}

\def\tng{\texttt{IllustrisTNG}}
\def\xmm{{XMM--Newton}}
\def\chandra{{Chandra}}
\def\eRosita{{eRosita}}
\usepackage{nicefrac}
\def\Mone{$M_1$}
\def\Mtwo{$M_2$}
\def\Mthree{$M_3$}
\def\Mfour{$M_4$}

\def\tng{IllustrisTNG}

\def\fgas{$f_{\mathrm{gas}}$}
\def\paperOne{Paper~I}
\makeatletter
\renewcommand*\aa@pageof{, page \thepage{} of \pageref*{LastPage}}
\makeatother

\title{X-ray emission in IllustrisTNG circumcluster environments} 
\subtitle{II. Possible origins of the soft X-ray excess emission}
\titlerunning{Soft X-ray emission in TNG}
\author{
  Celine~Gouin\inst{1}\thanks{E-mail:~\tt{gouin@iap.fr}},    
  Daniela~Gal\'arraga-Espinosa\inst{2, 3}, 
  Massimiliano~Bonamente\inst{4},
  Stephen~Walker\inst{4}, 
  Mohammad~Mirakhor \inst{4},
  Richard~Lieu \inst{4},
  Clotilde~Laigle\inst{1},
  Etienne Bonnassieux\inst{5},
  Charlotte Welker\inst{6},
  Stefano Gallo\inst{7},
  Tony Bonnaire\inst{8},
  Jade Paste \inst{1}
}

\institute{
Sorbonne Université, UMR7095, Institut d’Astrophysique de Paris, 98 bis Boulevard Arago, 75014 Paris, France
\label{inst1}
\and
Max-Planck Institute for Astrophysics, Karl-Schwarzschild-Str. 1, D-85741 Garching, Germany
\label{inst2}
\and
Kavli IPMU (WPI), UTIAS, The University of Tokyo, Kashiwa, Chiba 277-8583, Japan\label{IPMU}
\and
University of Alabama in Huntsville, Department of Physics and Astronomy, Huntsville, AL 35899, USA
\label{inst3}
\and
Instituto de Astrofísica de Andalucía, Consejo Superior de Investigaciones Científicas (CSIC), Glorieta de la Astronomía s/n, E-18008, Granada, Spain
\label{inst4}
\and
New York City College of Technology, City University of New York, Physics Department, Brooklyn, NY, USA
\label{inst5}
\and
Université Paris-Saclay, CNRS, Institut d’Astrophysique Spatiale, 91405, Orsay, France
\label{inst6}
\and
Laboratoire de Physique de l'Ecole Normale Sup\'erieure, ENS, Universit\'e PSL, CNRS, Sorbonne Universit\'e, Universit\'e Paris Cit\'e, F-75005 Paris, France
\label{inst7}
}

\date{\today}

\abstract
{
\textit{Context.} 
An excess of soft X--ray emission ($\sim 0.2-1$~keV) above the contribution from the hot intracluster medium (ICM) has been detected in a number of galaxy clusters, including the Coma cluster. The physical origin of this emitting medium above the hot ICM has not yet been determined, in particular, it is unclear whether it is thermal or nonthermal.\\
\textit{Aims.} 
We investigate the gas phase and gas structure that reproduce the soft excess radiation from the cluster core to the outskirts best using simulations.\\
\textit{Method.} 
By using the \tng\ simulation (TNG300), we predict the radial profile of thermodynamic properties and the soft X--ray surface brightness of 138 clusters within $5 \times r_{200}$.
Their X--ray emission was simulated for the hot ICM gas phase ($T \geq 10^7 \ K$), the entire warm--hot medium at a temperature $T=10^{5-7} K$ (WARM), and for the diffuse and low--density warm--hot intergalactic medium (WHIM).\\
\textit{Results.} 
The soft excess inside clusters appears to be produced by substructures of the WARM gas phase that host dense warm clumps, that is, the warm circumgalactic medium (WCGM), and the inner soft excess is strongly correlated with substructure and the WCGM mass fractions. Outside of the virial radius, the fraction of WHIM gas that is mostly inside filaments that are connected to clusters boosts the soft X--ray excess. The more diffuse the gas, the higher the soft X-ray excess beyond the virial region.  \\
\textit{Conclusion.} 
The thermal emission of the WARM gas phase in the form of WCGM clumps and WHIM diffuse filaments reproduces the soft excess emission that was observed up to the virial radius in Coma and in the inner regions of other massive clusters. Moreover, our analysis suggests that soft X--ray excess is a proxy of the dynamical cluster state and that higher excess is observed in the most unrelaxed clusters.}


\authorrunning{Gouin et al.}

\begin{document}
\maketitle

\section{Introduction}

A hot and diffuse intracluster medium (ICM) is the dominant reservoir of baryonic matter in galaxy clusters, with a typical mass fraction (\fgas) of 10-20\% of the total mass that is required to keep the hot gas in hydrostatic equilibrium with an unseen dark matter component \citep[e.g.][]{allen2004,landry2013, mantz2022}.
With typical temperatures of $\log_{10} T(\mathrm{K}) \geq 7$ \citep[e.g.][]{leccardi2008}, the hot ICM is readily observable with imaging X--ray detectors, especially at photon energies $\gtrsim 1$~keV, where the interstellar medium (ISM) Galactic absorption is less significant \citep[e.g.][]{dickey1990, kalberla2005,HI4PI2016}.

At large scales, different gas phases dominate different cosmic web environments, according to cosmological hydrodynamical simulations \citep{martizzi2019}. In general, the densest structures such as the cluster cores are mainly in the form of hot plasma. In contrast, filaments and sheets are mostly populated by warm gas, and voids are dominated by cold gas in the form of the intergalactic medium (IGM). Matter flows from filaments to clusters in the large-scale filamentary network \citep{Hahn2015}. In this context, hydrodynamical cosmological simulations are our best option for understanding the physics of this gas that is accreted from filaments to clusters.
Numerous physical processes are thus explored in state-of-the-art simulations, such as the turbulence induced by large cosmic flow \citep{VallesPerez2021, Mohapatra2022}, merging events and accretion shocks \citep[see e.g.][]{Shi2020}, the presence of cool-core \citep{Braspenning2023}, and the dependence on baryonic physic models, such as Active Galactic Nuclei (AGN) feedback, in pushing out gas from cluster centers \citep{Hahn2017}.

Numerous physical processes are thus explored in advanced simulations, such as the turbulence induced by large cosmic flow \citep{VallesPerez2021, Mohapatra2022}, merging events \citep[see e.g.][]{Shi2020}, and the presence of cool core \citep{Braspenning2023}. In addition, the dependence on baryonic physic models, such as Active Galactic Nuclei (AGN) feedback, is also affecting the gas distribution, in pushing out the gas from cluster centers \citep{Hahn2017}.
In particular, by using the recent results from eROSITA, \cite{Bahar2024} have recently attempted to constrain AGN feedback from the X-ray emission of groups. 
Recently, \cite{Soumya_model} have proposed a novel forward model to constrain the AGN X-ray luminosity jointly with the radial hot-gas distribution based mock X-ray emission from \tng\ simulation \citep{Soumya_LC}.

At soft X-ray energies ($\sim$ 0.1-1 keV), some clusters show an excess of soft X-ray emission within the virial radius. This is defined as an excess of counts above the expected soft X-ray emission from the hot ICM gas as modeled from its harder X-ray emissions.
This soft excess was first observed in Extreme Ultraviolet Explorer (EUVE)  observations of the Coma and Virgo clusters \citep{lieu1996a,lieu1996b,bowyer1996} and was later observed by {ROSAT} in different clusters \citep{bonamente2001,bonamente2001c,bonamente2002}.
This excess was also observed by X-ray Multi-Mirror Mission (\xmm) \ \citep[e.g.][]{nevalainen2003, kaastra2003, bonamente2005}, {Beppo-SAX} \citep{bonamente2001b}, and even by \chandra\ \citep{bonamente2011}. 
Measurements of the soft excess in particular rely on the accuracy of 
(a) the estimate of the hot ICM contribution to the soft band, 
(b)  the background subtraction, and 
(c) the effect of the foreground Galactic absorption, which varies in time and space \citep[e.g.][]{snowden1994,snowden1995}. 
Although some of the detections have been questioned or revised based on the uncertainties on these aspects of the data analysis \citep[e.g.][]{bowyer1999, nevalainen2007, bonamente2011}, the soft excess is now an established observational feature of some galaxy clusters. In particular, we focus on the comparison with the strong soft excess out to the virial radius for the Coma cluster, which is one of the most massive and nearby clusters \citep{bonamente2003, bonamente2009, mirakhor2020, bonamente2022c}.

The interpretation of this soft excess has so far been elusive because it has not been possible to confirm emission lines associated with the excess with a high statistical probability \citep[as reported by][]{kaastra2003, finoguenov2003} because the resolution of the X--ray instruments is limited.
It is unclear whether the soft excess originates from within the virial cluster radius \citep[e.g., as originally investigated by][]{cheng2005} or from external filaments that are projected against the cluster \citep[e.g., as investigated in][]{churazov2023}. It is therefore of interest to study the properties of the baryonic phases within the virial radius, where the hot gas is the dominant baryonic form, and in its immediate vicinity, where warm subvirial gas is still accreting toward clusters and is expected to be densest and therefore more readily observable.

In \cite{gouin2023} (hereafter \paperOne), we began to investigate the gas properties and soft X-ray emission in the circumcluster environment, which is defined as the region within a few virial radii of galaxy clusters, using the \tng\ simulation \citep{TNG,TNG2}. In this follow-up paper, we focus on the origin of the soft X--ray emission and on its dependence on the gas properties. 
This paper is structured as follows. We describe
in Sect.~\ref{sec:methods} the numerical methods we used to predict soft X--ray emission in the circumcluster environment. 
In Sect.~\ref{sec:comparison} we compare the soft X--ray predictions to current observations and discuss the detectability of warm gas.
In Sect.~\ref{sec:Origin} we investigate the possible origin of this soft excess as a function of the gas properties and gas structures. In Sect.~\ref{sec:Disc} we discuss our results in the context of recent observations and predictions, and we present our conclusions in Sect.~\ref{sec:conclus}.

\section{Numerical predictions of the soft X--ray emission \label{sec:methods}}

The \tng\ simulation and the analysis methods used for this project were presented in \cite{gouin2023}, and they are briefly summarized here. 
We used the TNG300-1 simulation at $z = 0$, which is the largest volume ($\sim 300 \ \textrm{Mpc}^3$) in the \tng\ suite with the highest spatial resolution. This provides a statistically significant cluster sample and enabled us to model the baryonic physics in detail \citep{TNG,TNG2}.
We divided the baryons into different phases as defined in Table~\ref{tab:gas}: the hot ICM at $\log T(\mathrm{K}) \geq 7$, and the warm gas at $5 \leq \log T(\mathrm{K}) \leq 7$ (WARM), which is further subdivided into the denser warm--hot circumgalactic medium (WCGM) at $n_e \geq 10^{-4}$~cm$^{-3}$ and the diffuser warm--hot intergalactic medium (WHIM) at lower density, which is typically found in filaments \citep[see e.g.][]{galarraga2021,gouin2022}. Thermodynamic quantities of interest such as the density, temperature, chemical abundance, and clumpiness of the gas were averaged in spherical shells out to $5 \times r_{200}$, which is the outermost radius we considered.
As explained in \paperOne, we did not compute the thermodynamic profiles (and thus, the soft X--ray emission) of the WCGM gas alone because there are not enough particles from which we might obtain smooth profiles because their distribution fluctuates strongly. 
Instead of the WCGM phase alone, we combined the WCGM and WHIM gas into the WARM gas phase to explore the 
soft X--ray emission of the diffuse (WHIM) and clumpy (WCGM) warm gas, in contrast to the hot ICM.
Our sample of 138 clusters was divided into three subsets ordered by increasing mass, as shown in Table~\ref{tab:sample}, as we did in \paperOne.

\begin{table*}[]
    \caption{Definition of the gas phases. The WARM gas is the sum of the WHIM and WCGM gas phases.}
    \centering
    \begin{tabular}{c| l| l| l }
         Gas phase & Density [$cm^{-3}$] & Temperature [$K$]  & Environments \\[2pt]
        \hline 
         HOT gas  &  - & $T>10^7$ &  Dominant inside cluster ($r<r_{200}$), referred as hot ICM  \\[2pt]
        \hline 
         WARM gas & - & $10^5<T<10^7$ & Dominant outside cluster ($r>r_{200}$) \\[2pt]
         \hdashline
         WHIM gas & $n_e < 10^{-4}$~cm$^{-3}$ & $10^5<T<10^7$ & Dominant inside filaments \citep{galarraga2021} \\[2pt] 
          \hdashline
         WCGM gas & $n_e \geq 10^{-4}$~cm$^{-3}$ & $10^5<T<10^7$ & Clumps of gas tracing filaments \citep{angelinelli2021} \\[2pt] 
    \hline
    \end{tabular}
    \label{tab:gas}
\end{table*}

\begin{table*}[]
    \caption{Main properties of the entire cluster sample and its four mass bins $M_1$, $M_2$, $M_3$, and $M_4$. Sample $M_4$ is also included in $M_3$. Sample $M_4$ is composed of the five most massive clusters and represents a Coma-like cluster sample.}
    \centering
    \begin{tabular}{c| l| l| l | l}
         Sample & Mass Range & $\langle M_{200}\rangle$  & $\langle r_{200}\rangle$ & Number of clusters \\[2pt]
        \hline 
         All clusters & $\log( M_{200}[M_{\odot}/h])  >14$ & $1.97 \times 10^{14} M_{\odot}/h$ & 0.92/$h$ Mpc& 138 \\[2pt]
         $M_1$ & $14.0 <\log( M_{200}[M_{\odot}/h]) <14.1$ & $1.13 \times 10^{14} M_{\odot}/h$ & 0.78/$h$ Mpc & 47\\[2pt]
         $M_2$ & $14.1 <\log( M_{200}[M_{\odot}/h]) <14.5$& $1.91 \times 10^{14} M_{\odot}/h$ & 0.92/$h$ Mpc& 78\\[2pt]
         $M_3$ & $ \log( M_{200}[M_{\odot}/h]) >14.5$ & $5.37 \times 10^{14} M_{\odot}/h$ & 1.3/$h$ Mpc & 13\\[2pt]
         \hdashline[0.5pt/5pt]
         $M_4$ & $ \log( M_{200}[M_{\odot}/h]) >14.75$ & $7.61 \times 10^{14} M_{\odot}/h$ & 1.5/$h$ Mpc & 5\\
    \hline
    \end{tabular}
    \label{tab:sample}
\end{table*}

\subsection{Thermodynamic quantities}
\label{sec:thermo}

We begin by summarizing the gas profiles for the two dominant gas phases in the cluster environment, the hot ICM and the WARM gas.
These two phases are necessary to simulate the X-ray emission because they are the two dominant gas phases in the circumcluster environments \citep{martizzi2019, gouin2022}. 
The method for computing the radial profiles of the thermodynamic properties is detailed in \paperOne, and it is summarized here. For each cluster, we computed the radial profiles of the thermodynamic quantities within spherical shells centered on the cluster, and the cluster-centric distance was normalized by $r_{200}$. Each profile was obtained by averaging over the gas cells of a given phase using a volume-weighted average for each gas cell.
For the temperature, we tested three different weights, the cell volume, the mass, and the emissivity -weighted average. We concluded that they do not affect our results significantly. Moreover, we corrected the metallically profile by a factor $1.6$ following \cite{Vogelsberger2018}. These authors found that  the TNG300-1 metallicity profiles should be shifted by a factor of 1.6 to match the highly resolved TNG100-1 simulations. This discrepancy was also found in Cluster-EAGLE clusters by \cite{Barnes2017}.

\begin{figure*}
   \centering
    \includegraphics[width=8.8cm]{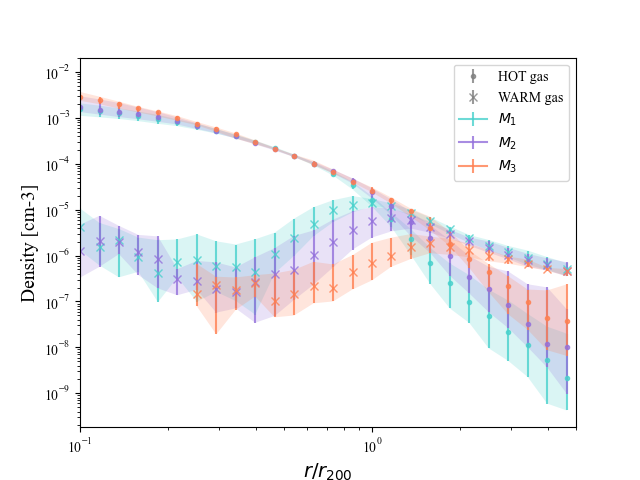} 
    \includegraphics[width=8.8cm]{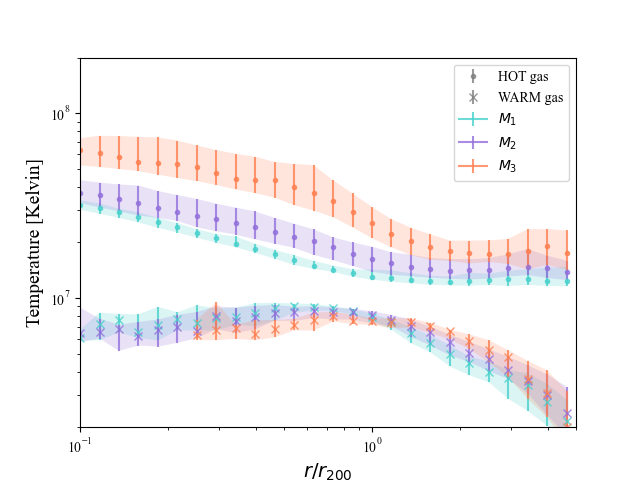} \\
    \includegraphics[width=8.8cm]{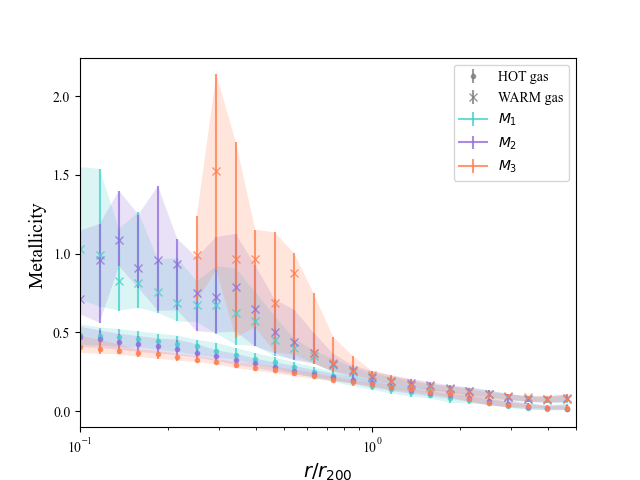}
    \includegraphics[width=8.8cm]{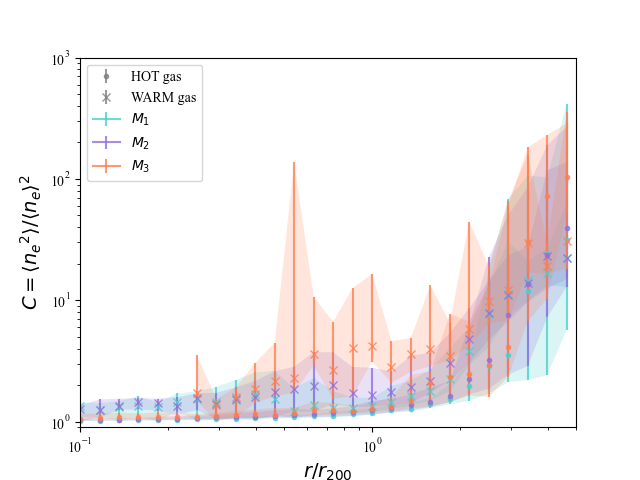} 
    \caption{Median radial profile of the gas properties for the two main gas phases (HOT and WARM gas, as defined in Table~\ref{tab:gas}) as a function of the mass bins $M_1$, $M_2$, and $M_3$, plotted in blue, purple, and orange, respectively (as defined in Table~\ref{tab:sample}). The error bars are the percentiles from 20\% to 80\%. The average density (top left), temperature (top right), metal abundance (bottom left), and clumpiness (bottom right) profiles are illustrated in the different panels.}
    \label{fig:THERMO}
\end{figure*}

Figure~\ref{fig:THERMO} illustrates the averaged radial profiles of the gas density, temperature, and metal abundance for our three main cluster mass bins.
The top left panel displays the mean density within spherical radial shells centered on the clusters. We found that the radial density of the WARM gas is significantly affected by the cluster mass. Compared to high-mass clusters, low-mass clusters contain denser WARM gas within approximately the virial radius, and the behavior beyond it is different. 
The radius at which the WARM density becomes higher than the HOT gas density occurs at about $r=0.9 \ r_{200}$ for $M_1$, at $1.3 \ r_{200}$ for $M_2$, and at $2.1 \ r_{200}$ for $M_3$. This shows that the hot ICM dominates the cluster region out to larger radii for more massive clusters.
The top right panel shows that the WARM temperature does not depend on the cluster mass, unlike for the hot ICM gas. As expected and well established observationally, clusters with higher masses have hotter ICM temperatures than lower mass clusters because their gravitational wells are deeper. This also naturally accounts for the faster transition from HOT to WARM in lower-mass clusters. 
The metal abundance in the WARM gas in the bottom left panel is significantly higher than in the ICM. The radial gradient starts from near solar levels at the center and decreases to approximately 10\% solar at the virial radius. Beyond this radius, the abundance in the WARM phase is subsolar. The metallicity profile agrees with findings from the TNG-Cluster Simulation \citep{Nelson2023}, which indicated a minor sensitivity to the halo mass and achieved about $0.4 \ Z_{\odot}$ in the cluster centers.
Comparisons between our simulated density, temperature, and metallicity profiles and the observations and other simulations were presented in \paperOne. The results agreed well. \\

The X--ray surface brightness is  proportional to the square of the density. It is thus highly sensitive to inhomogeneities in the emitting phases. The radial profiles of the clumpiness factor of the various phases is defined as
\begin{equation}
    C = \dfrac{\langle n^2 \rangle}{\langle n \rangle^2} \,,
\end{equation}
with $n$ the gas density \citep[see][for a definition]{nagai2011}.
The bottom right panel of Fig.~\ref{fig:THERMO} shows the clumpiness profile of the HOT and WARM gas. The WARM gas is substantially more clumpy than the hot ICM within and beyond the virial radius and for all cluster mass bins. Moreover, the clumping of the WARM gas and the cluster mass beyond $0.5 \times r_{200}$ is strongly correlated, such that more massive clusters have clumpier WARM gas around them. The clumpiness of the gas plays a crucial role when the subsequent X--ray emission is investigated. We show below that the clumpier the WARM gas, the higher the soft X--ray excess.
Equivalent clumpiness measurements, along with comparisons between our analysis, observations, and other simulations, were also presented in \paperOne.

\subsection{Masses and gas-mass fractions}
\label{sec:mass}

\begin{figure}
   \includegraphics[width=3.5in]{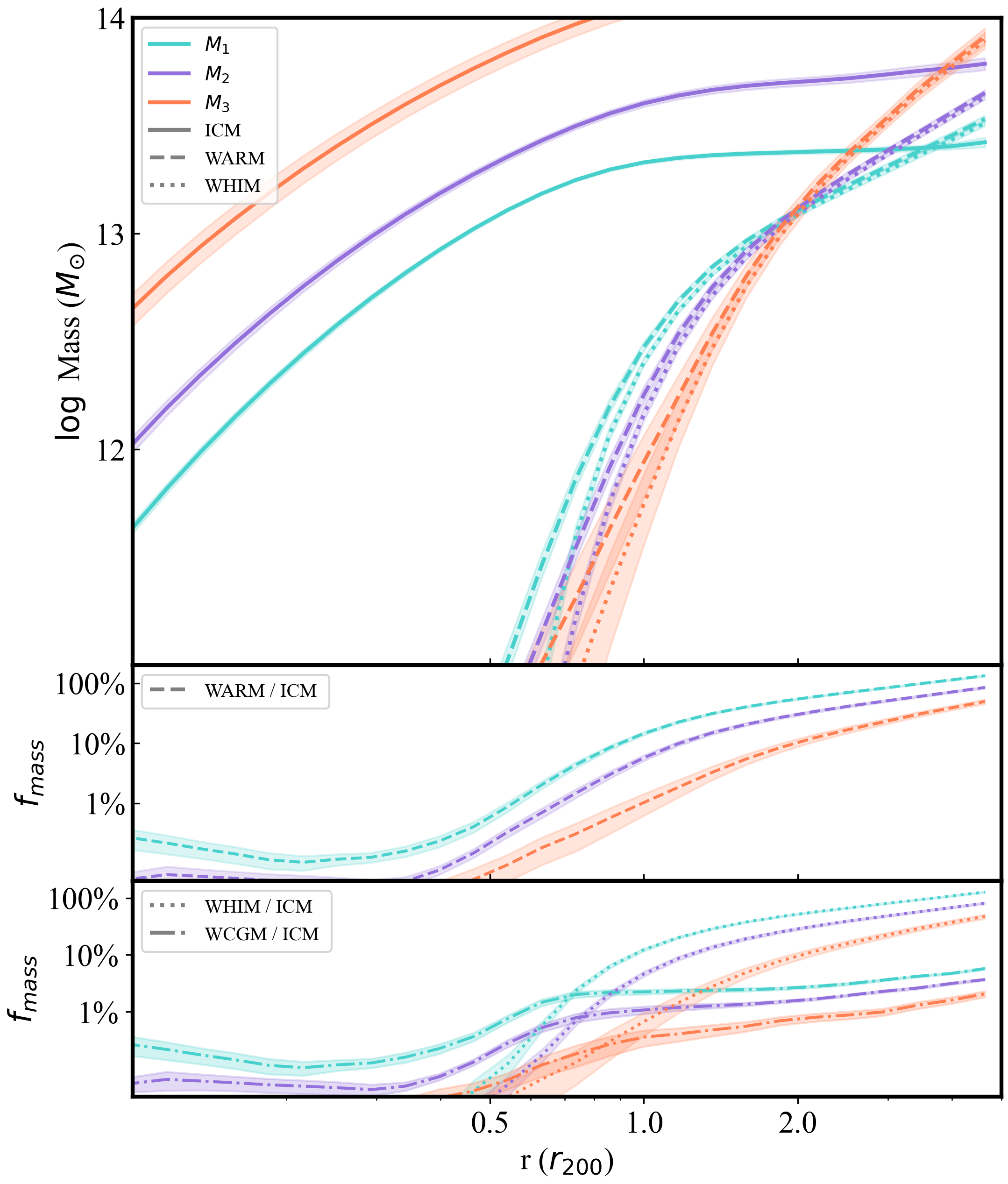}
    \caption{Top panel: Mean radial profiles of the integrated ICM (solid lines), WARM (dashed lines), and WHIM (dotted lines) gas masses for the three mass bins $M_1$(salmon lines), $M_2$ (purple lines), and $M_3$ (blue lines).
    Bottom panels: Radial profile of the mass fraction of WARM gas (second panel), composed of WHIM and WCGM (third panel), in the hot ICM mass. The error bars are the error on the mean.}
    \label{fig:MASS}
\end{figure}

The mass of the various gas phases was obtained as a straightforward integration of the density profiles of Fig.~\ref{fig:THERMO}. An estimate of the total mass implied by the assumption of hydrostatic equilibrium (HSE) can be also calculated (as shown in \paperOne). In this paper, we instead focus on the relative ratios of the baryonic masses, and we do not discuss hydrostatic masses.
 
Figure~\ref{fig:MASS} shows the radial profiles of the integrated ICM, WARM, and WHIM gas masses (top panel) as a function of cluster mass. 
In the bottom panels, we illustrate the mass ratios of WARM to HOT,  WHIM to HOT, and WCGM to HOT, computed as integrated gas-mass profiles. 
This analysis of the \tng\ simulation provides a number of useful clues as to the locations of baryons in circumcluster environments. We list these clues below.
\begin{itemize}
    \item In the inner cluster regions ($r \leq 0.5 \times r_{200}$), only lower--mass clusters have a significant amount of WARM gas. It is at a level of $\sim 0.1-1$~\% of the hot ICM. This WARM gas is in the WCGM gas phase, as expected previously from the high density of the WARM gas inside these low-mass clusters.
    \item Near the virial radius, most of the WARM gas in lower--mass clusters is in the form of diffuse WHIM gas, while high--mass clusters have comparable WCGM and WHIM gas masses. This is expected from the radial extension in the ICM of higher--mass clusters, which are located relatively farther away from their center than lower--mass clusters. 
    \item  Interestingly, clusters appear to have a constant radius of about $2\times r_{200}$ whithin which the WARM gas mass is approximately $2 \times 10^{13}$~M$_{\odot}$, regardless of the cluster mass. Beyond this radius, the WARM mass increases in proportion to the cluster mass, as expected. At these large radii ($r>2 \times r_{200}$),  almost all the WARM gas is in the form of diffuse WHIM, whereas the WCGM clumps represent only about $~5\%$ of the WARM gas mass.
\end{itemize}

\subsection{The X--ray surface brightness}
\label{sec:Sx}

The X--ray surface brightness is defined as 
\begin{equation}
S_X(\theta)= \dfrac{1}{4 \pi (1+z)^4}\int_l n_e n_H \Lambda(T,A)\, dl(\theta),
\label{eq:Sx}
\end{equation}
and it is given in units of [erg~cm$^{-2}$~s$^{-1}$~Srad$^{-1}$]. The radiation is assumed to be isotropic by the $4 \pi$ factor (see Eq.~5 of \paperOne).
In Eq.~\ref{eq:Sx}, $\Lambda(T,A)$ is the emissivity of the plasma, calculated in [keV~s$^{-1}$~cm$^3$] with the software \texttt{pyAtomDB}, as shown in \paperOne, and it is a function of the average plasma temperature and abundance. $n_e$ and $n_H$ are the electron and hydrogen number density in [cm$^{-3}$], respectively, and the integration
is made along the sightline at a give projected radial distance $\theta$.
The X--ray surface brightness is also known as the intensity of the radiation $I$, or of the energy emitted per unit time,  solid angle, and detector area perpendicular to the radiation. The intensity is a convenient astrophysical quantity because it is a quantity that is (nearly) independent of distance or redshift because the solid angle of the emitted radiation and the surface area of the detector scale in opposite ways with the distance between the detector and the source \citep[e.g.][]{rybicki1979}.
 Because the radiation is optically thin, the surface brightness was obtained as an integration along the sightline at a constant projected angular distance $\theta$ of $n_e n_H \Lambda(T,A)$ in 50 radial bins between 0.05 and $5\times r_{200}$, which correspond to the same number of cylindrical shells, as described in detail in \paperOne, where details of the projection of the 3D radial profiles (e.g., Fig.~\ref{fig:THERMO}) were discussed. In particular, the use of these azymuthally symmetric 3D radial profiles means that the effect of triaxial geometry is accounted for in the projection in the sample sense because there is no choice of a specific axis over which the projection is made.

\begin{figure}
    \centering
    \includegraphics[width=3.5in]{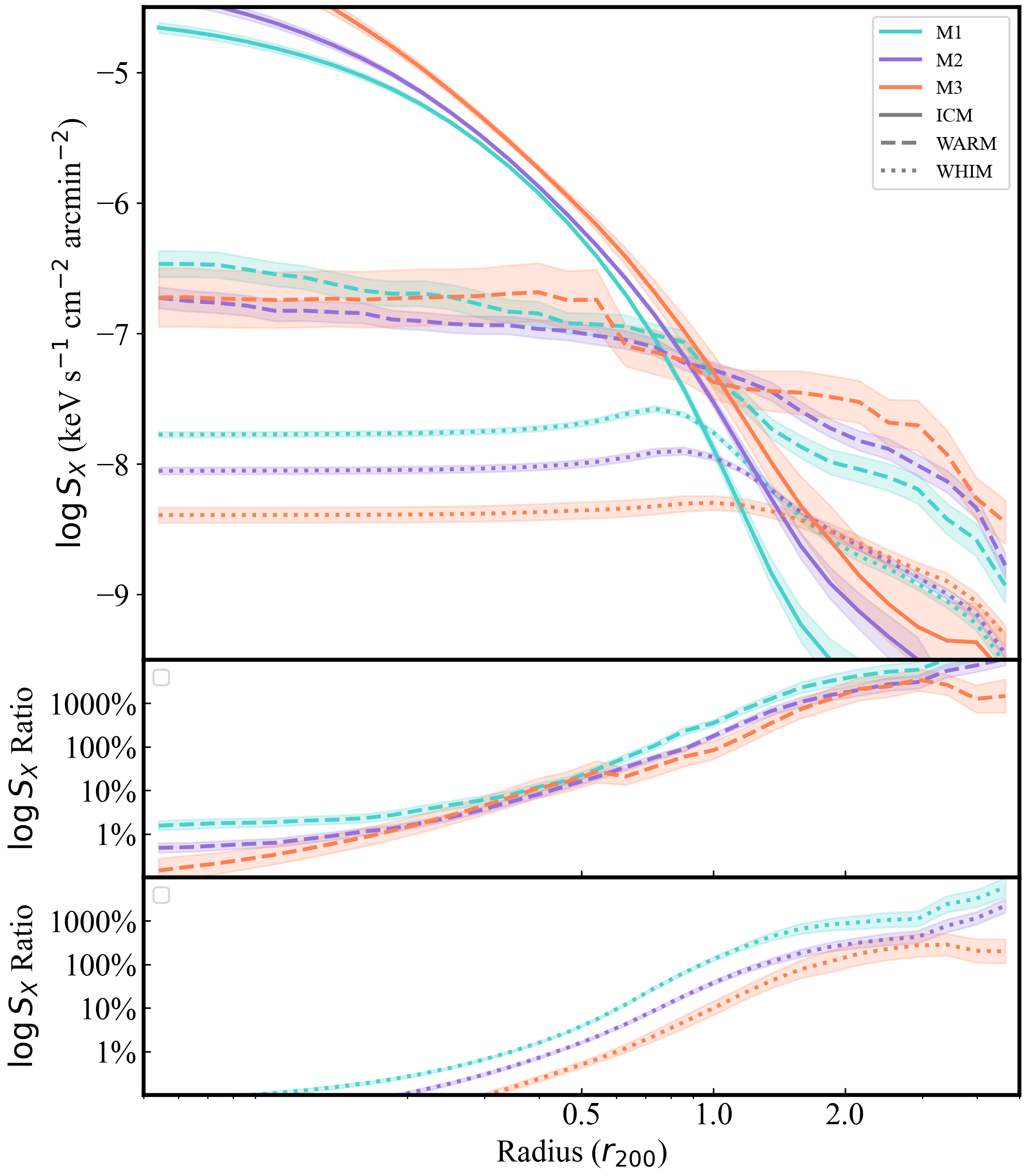}
    \caption{Top panel: Mean radial X--ray surface brightness profiles for the three cluster mass bins (\Mone, \Mtwo, and \Mthree) simulated in the 0.2-0.4~KeV band. 
    Bottom panels: WARM–to-ICM (dashed lines) and WHIM–to-ICM (dotted lines) surface brightness ratios. The error bars are the error on the mean.}
    \label{fig:SxMassPlot}
\end{figure}

Considering this formalism, we used the thermodynamic radial profiles of each cluster to predict its associated X--ray surface brightness profile. Figure~\ref{fig:SxMassPlot} shows the average surface brightness profiles for our three cluster mass ranges considering an X--ray band of 0.2-0.4~keV. 
The results and their interpretation are very similar for the second soft X--ray band of interest (0.2-1~keV), and this band is not shown in the figure.  
We highlight the relative importance of the X--ray surface brightness from the different gas phases (ICM, WARM, and WHIM) as a function of the cluster--centric distance. 
This emission ratio of the HOT and WARM gas must be consistent with the observed soft X-ray excess, which is defined as the excess counts above the expected hot ICM emission. It is compared to recent observations in Sect.~\ref{sec:comparison}.
The hot ICM dominates the soft X--ray emission up to approximately the virial radius ($< 1 \ r_{200}$) for all mass ranges, as expected. 
To highlight the importance of other gas phases for the soft-X-ray emission, the lower panels of Figure~\ref{fig:SxMassPlot} plot the WARM--to-ICM and WHIM--to-ICM surface brightness ratios for the three mass samples.
Based on these radial profiles, we interpret the main features of the soft X--ray emission from the different gas phases below.
\begin{itemize}
    \item At all radii and for all mass ranges, the WCGM gas in the WARM phase apparently systematically boosts the soft-X emission because its density is higher. In contrast, 
    the diffuse WHIM gas alone produces less soft-X ray emission per unit mass.
    \item Within the virial radius ($ 1 \times r_{200}$), the dominant source of soft X--ray emission is the hot ICM. The second main contribution from the subvirial gas comes from the high--density WCGM, and the diffuse WHIM provides almost no emission up to $0.5 \times r_{200}$. This is true regardless of the cluster mass. 
    By comparing the WARM--to-ICM and WHIM--to-ICM surface brightness ratios, we found that the contribution from WARM gas contribution is far higher inside the clusters than that of the WHIM ($< 1 \times r_{200}$).  
    This strong radial evolution of the WARM--to--ICM surface brightness ratio, which reaches $100 \%$ at $1 \times r_{200}$, arises from the rapid increase in the WARM gas mass inside the clusters, as illustrated in Fig.~\ref{fig:MASS}.
    \item At the periphery of the cluster and beyond ($> 1 \times r_{200}$), the WARM gas predominantly contributes to the soft X--ray surface brightness. Notably, in this radial range, the higher the cluster mass, the more pronounced the emission from the WARM gas. 
     The analysis of the WARM-to-ICM and WHIM-to-ICM surface brightness ratios at these cluster distances reveals that the low-density WHIM gas primarily accounts for the soft X-ray excess, and that the contributions from the high-density WCGM serve as an enhancement. As highlighted in Fig.~\ref{fig:MASS}, the WCGM gas mass is considerably lower than that of the WHIM gas at these radii.
\end{itemize}

\section{Comparison between \tng\ predictions and soft X--ray observations \label{sec:comparison}}

This section compares observations of the soft X--ray emission and our predictions from \tng\ simulation. The X--ray surface brightness has a large cluster--to--cluster scatter beyond the mass dependence. We therefore focus on the relation between relevant gas properties (fraction of the gas phase and location in the subhalos and filaments) and soft X--ray excess in more detail in Sec.~\ref{sec:Origin} after we present the main comparison
between simulations and observations in this section.

\subsection{Comparison with the soft-X-ray emission profiles}

\begin{figure}
    \centering
    \includegraphics[width=3.5in]{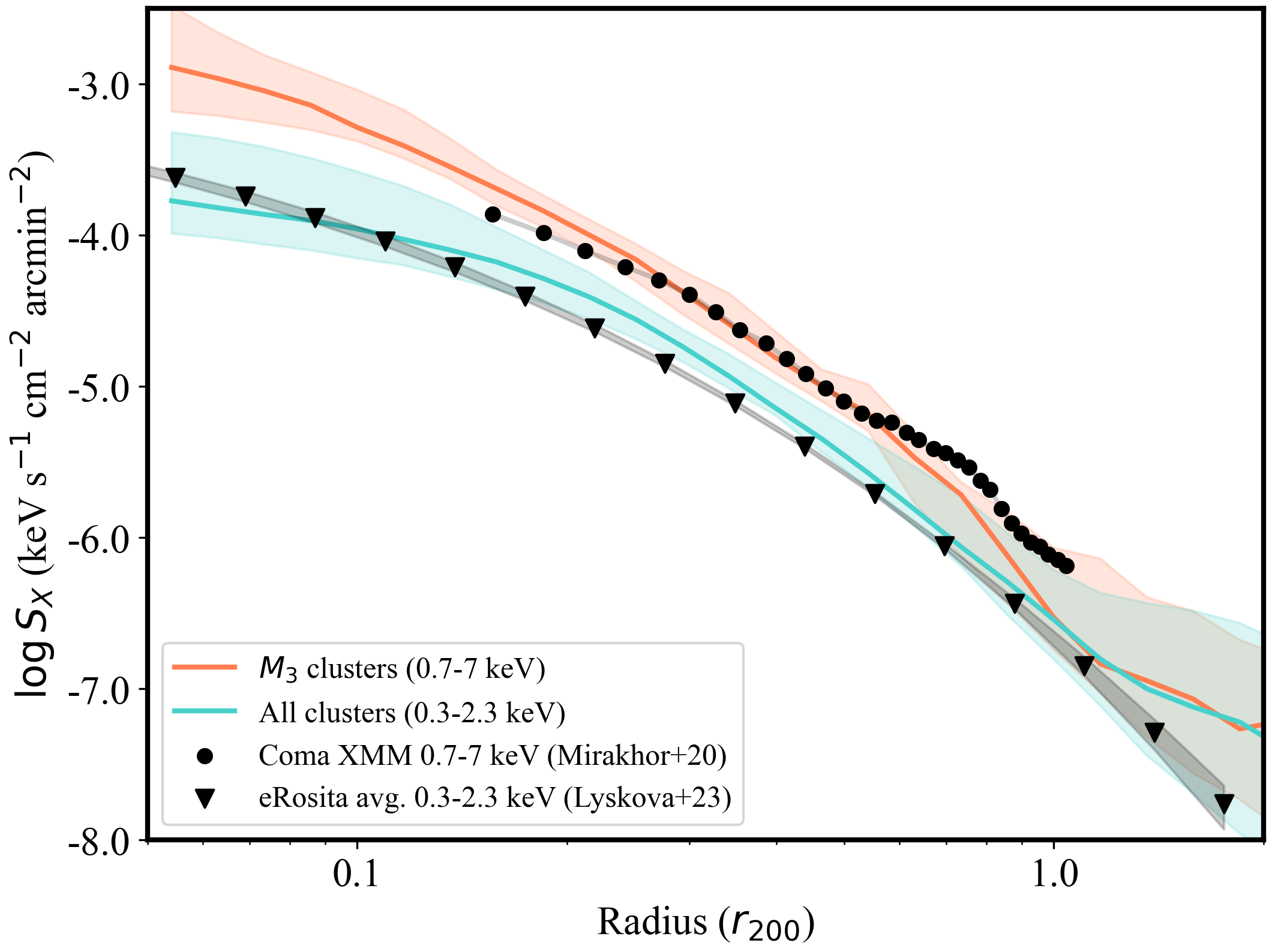}
    \caption{Median X--ray surface brightness profiles from \Mone\ and \Mthree~ clusters in the main--band X--ray ranges. The 0.3-2.3~keV range to match the best--fit model of the eRosita observations by \cite{lyskova2023} is shown in green, and the band of 0.7-7 keV to match the observations of Coma \citep{mirakhor2020} is shown as a blue curve. 
    The error bars in the simulated profiles are the percentiles from 20\% to 80\%.}
    \label{fig:SxPlot}
\end{figure}

Figure~\ref{fig:SxPlot} compares the total X--ray surface brightness (i.e., adding contributions from the hot-ICM and the WARM gas) for simulated cluster samples in the X--ray band of interest with selected observational results in main--band soft X--rays ($0.3 - 2.3$ keV for eRosita, and $0.7 - 7$ keV for XMM Newton).
As a first comparison, Figure~\ref{fig:SxPlot} illustrates the overall reasonable agreement between our analysis of the \tng\ simulation and X--ray observations of the X--ray surface brightness. The soft X-ray emission of the massive Coma cluster measured by \cite{mirakhor2020} in the $0.7$-$7$ keV band, illustrated as circles, matches the largest population of massive \tng\ clusters well (mass range $M_3$; orange curve). Moreover, the averaged surface brightness profile of eRosita clusters measured by \cite{lyskova2023}, as triangles, agrees well with the surface brightness profiles from our full sample of $138$ simulated cluster groups. 
The two comparisons thus indicate that the \tng\ simulation is consistent in general with recent available X--ray observations, although the core slope can be affected when clusters have a cool core \citep[e.g.][]{arnaud2002,Braspenning2023}, and the external slope of the profile is probably affected by the environment of the cosmic web, such as gas filaments and clumps, as investigated further below.

\subsection{Comparison with observations of the soft X--ray excess}

We used the analysis of the soft X--ray excess described in Sec.~\ref{sec:Sx} to compare the \tng\ predictions with cluster observations. The soft excess emission is usually reported as a fractional excess of counts in a given annular region above the contribution from the hot ICM. This is equivalent to the ratio of the soft X--ray surface brightness of the HOT and WARM phases shown in Fig.~\ref{fig:SxMassPlot}.

\begin{figure*}
    \centering
    \includegraphics[width=3.5in]{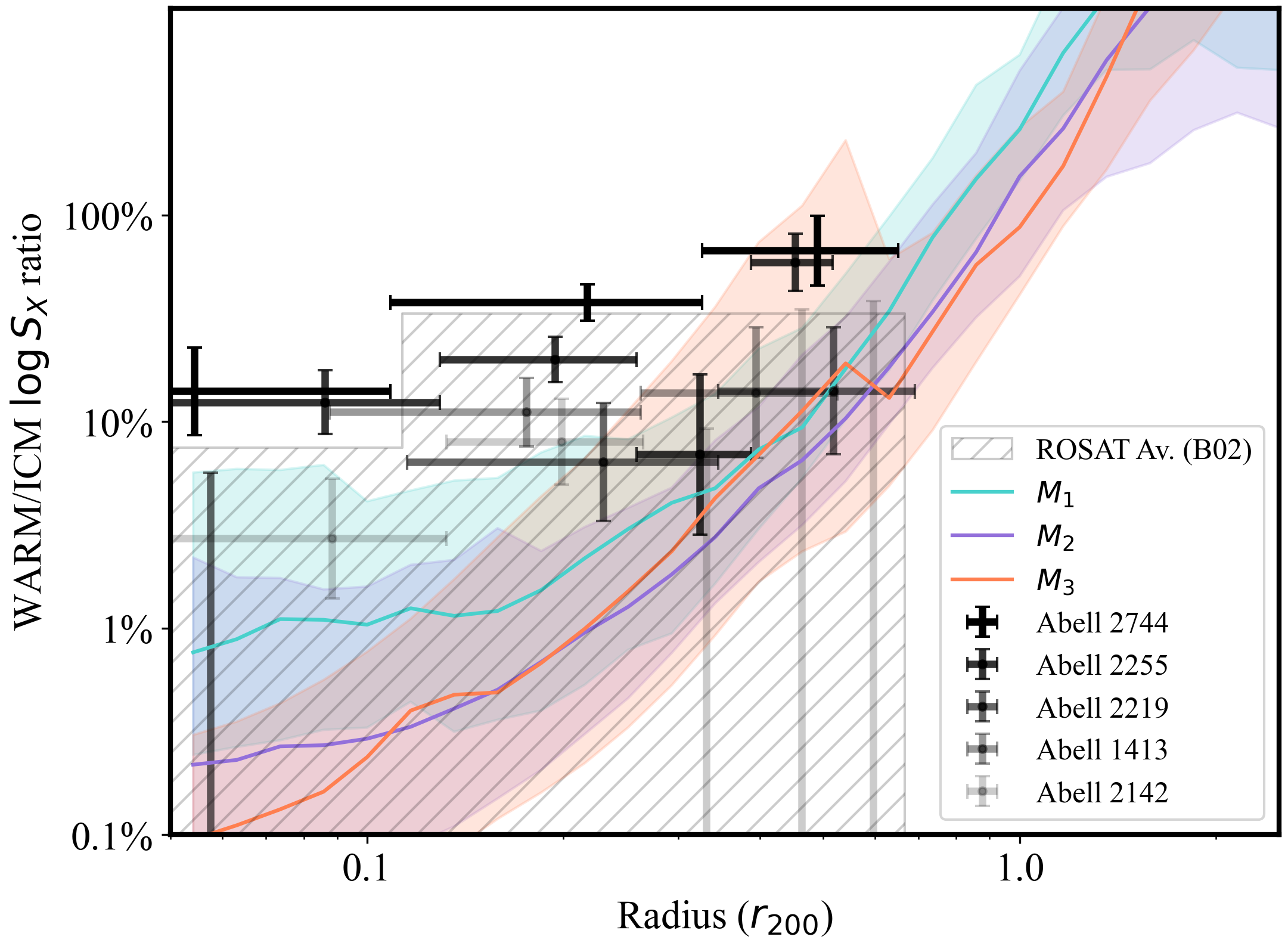}
    \includegraphics[width=3.5in]{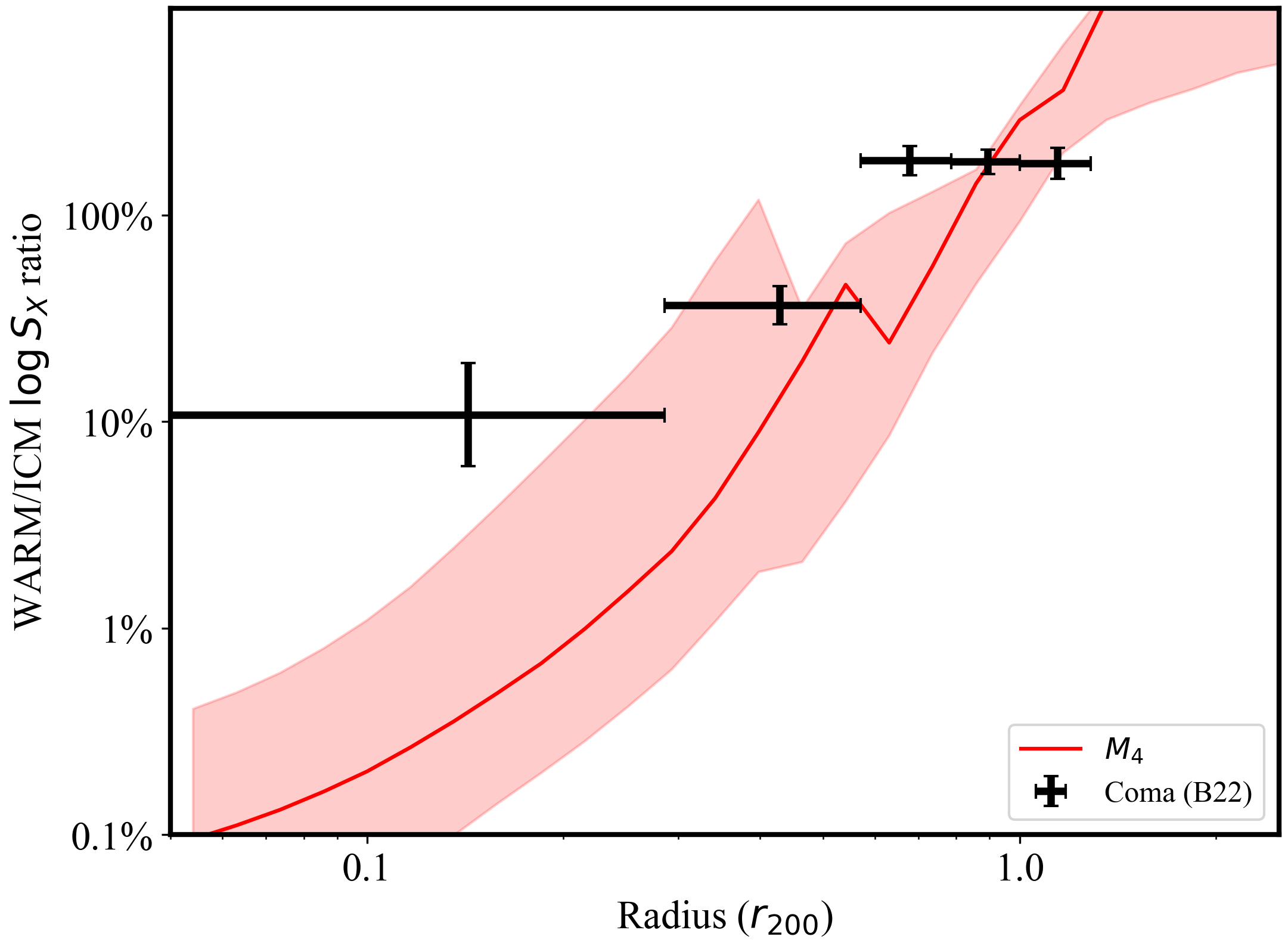}
    \caption{Left panel: Median WARM--to--ICM surface brightness ratio in the 0.2-0.4~keV band for the three cluster mass bins and the \rosat\ results of \cite{bonamente2002} for a sample of 38 clusters. 
    Right panel: Median WARM--to--ICM surface brightness ratio in the 0.2-1~keV band for clusters in the \Mfour\ mass bins and the observational results of the soft-X-ray excess from the Coma cluster \citep{bonamente2022c}. 
    For each line, the shaded area represents the 20th to 80th percentiles of the soft excess within each cluster sample. }
    \label{fig:ROSAT}
\end{figure*}

The left panel of Figure~\ref{fig:ROSAT} compares the \tng\ predictions with the \cite{bonamente2002} (hereafter B02) observations of the 0.2-0.4~keV band soft excess in a large sample of \rosat\ clusters.
The soft excess emission of about 18 of the 38 clusters observed with \rosat is statistically significant. The mean and standard deviation of the excess in the entire sample is shown as a shaded area in two broad radial bins: $r \leq 0.17$~Mpc with an excess of $2.4\pm5.7$~\%, and $0.17 \leq r \leq 1$~Mpc with an excess of $13\pm20$~\%, converted into the approximate $r/r_{200}$ in the figure. 
The specific data points of the 5 clusters with the largest radial coverage in the sample are also shown individually to highlight the typical detection significance.
The typical predicted soft excess in \tng\ in the 0.2-0.4~keV band is at the level of 0.1 to a few dozen percent, which agrees in general with the overall ROSAT observations.
Figure~\ref{fig:ROSAT} also compares the soft excess predictions between the \Mthree, \Mtwo\ and \Mone\ simulated samples. In particular, the lower--mass samples (\Mtwo\ and \Mone) predict a relatively higher central soft excess of up to 10\%, which may be more consistent with certain observed clusters, such as Abell~2744 and Abell~2255. 
Our result suggests that the soft excess in the core of our simulated clusters might be slightly lower than in observations on average (for $r \lesssim 0.2 \ r_{200}$). This might be explained by the simulation resolution (approximately a few kiloparsec), unresolved clumps, and by the fact that we did not account for nonthermal processes (see the discussion in Sect. \ref{sec:Disc}).

The right panel of Figure~\ref{fig:ROSAT} compares the \cite{bonamente2022c} (hereafter B22) \rosat\ observations of Coma out to the virial radius in the broader 0.2-1 keV band to the corresponding \tng\ predictions. 
The agreement is very good at all radii, especially near the virial radius, where \rosat\ measured a $\geq 100$~\% soft excess. This agrees with the \Mfour\ sample, which comprises the five more Coma--like massive clusters in \tng.
It is useful to remark that the B22 results are a combination of the \cite{bonamente2003} \rosat\ fluxes with a reanalysis of the hot ICM that is based on the more recent \xmm\ observations of Coma by \cite{mirakhor2020} in order to refine the hot ICM predictions in the soft band.

\subsection{Detectability of the soft-X-ray emission \label{sec:detectability}}

\begin{figure*}
    \centering
    \includegraphics[width=3.5in]{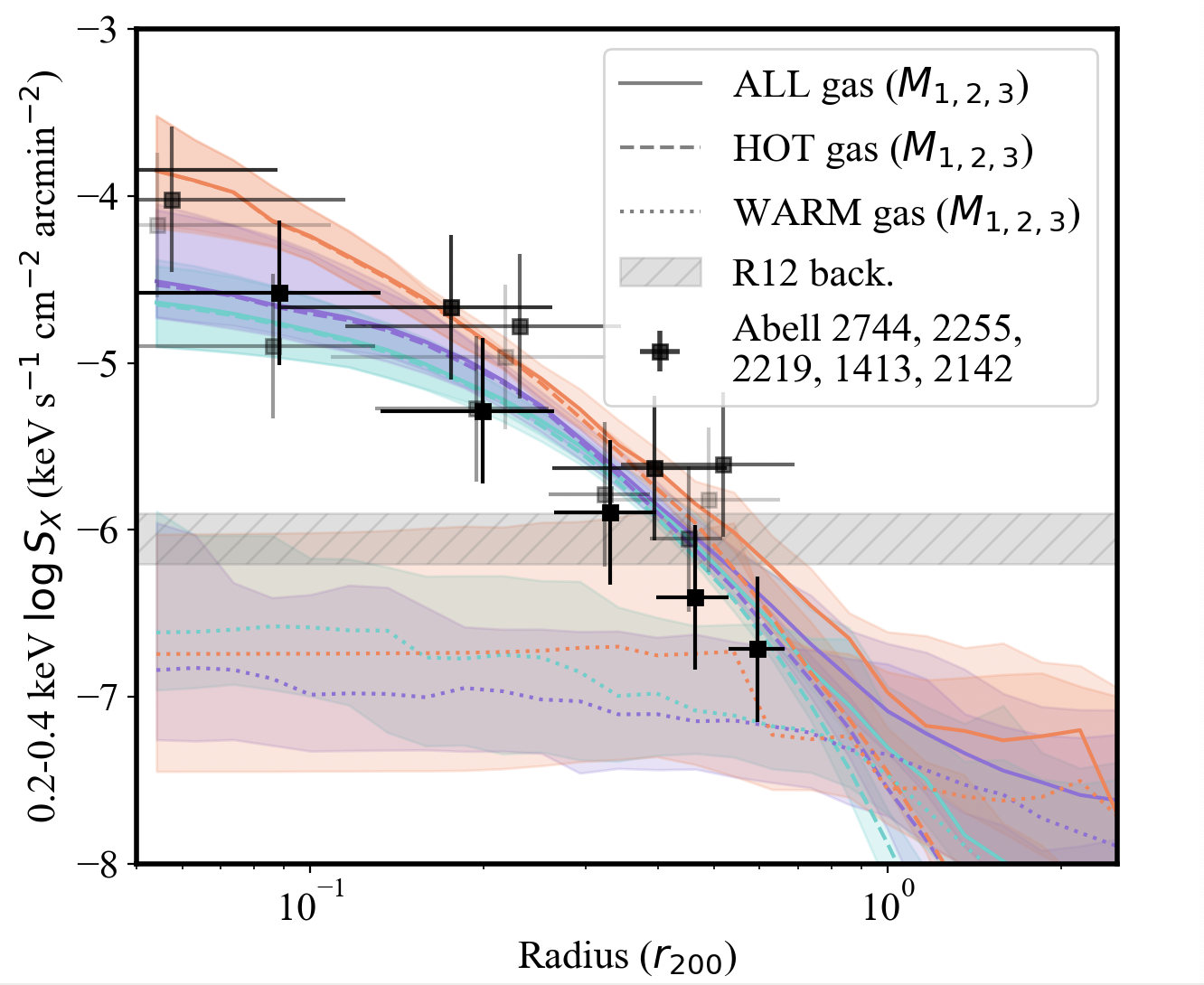}
    \includegraphics[width=3.5in]{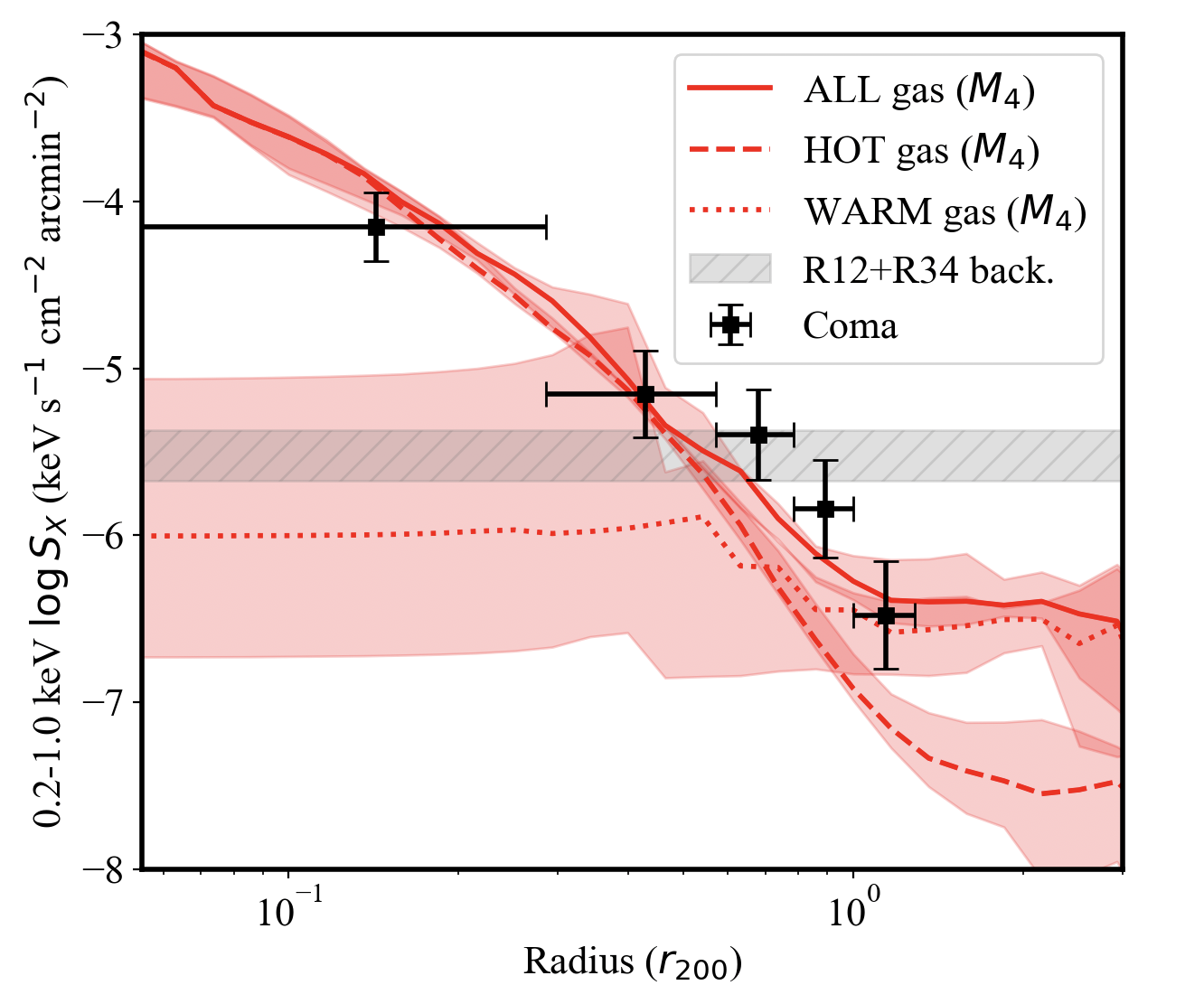}
    \caption{Left panel: Median soft X--ray surface brightness of the WARM and ICM phases and the ALL gas in the 0.2-0.4~keV band, approximately equivalent to the \rosat\ R12 band of the B02 data. Right panel:  Median soft X--ray surface brightness in the 0.2-1~keV band, corresponding to the B22 data. For each line, the shaded area represents the 20th to 80th percentiles within each cluster sample. The R12 band corresponds to approximately $0.2-0.4$~keV, and the R34 band corresponds to $\sim 0.4-1.0$~keV \citep{snowden1995, snowden1997}. }
    \label{fig:SxDetectability}
\end{figure*}

We complement our analysis of the soft X--ray excess emission in the circumcluster environment by providing an outlook on the detectability of the soft excess according to the \tng\ predictions.
The previous section has already shown that the available soft excess measurements are clearly consistent with our predictions. This section provides additional considerations, including the possibility of future detections near the virial radius, which are currently limited to the \cite{bonamente2003} and \cite{bonamente2022c} \rosat\ detections in the Coma cluster.

Figure~\ref{fig:SxDetectability} shows the soft X--ray surface brightness of the HOT, WARM, and ALL gas in the two bands of interest in order to assess whether the detection of the warm subvirial gas near clusters is possible. For comparison, we also show the average surface brightness of the soft X--ray sky as measured by \rosat\ \citep{snowden1995} as a measure of the typical intrinsic sky background level that is expected to be present at these wavelengths. The error bars in the simulated profiles represent the percentiles from 20\% to 80\%. 
We approximated the R12 \rosat\ band with  the 0.2-0.4~keV band, and we used an average effective area of the \rosat\ position--sensitive proportional counter (PSPC) of 120~cm$^2$ \citep[see, e.g., Fig.~1 of][]{snowden1994}.
The diffuse background at energies near $\nicefrac{1}{4}$~keV is highly variable in time and space, and we considered an average background surface brightness of 300--600~$\times 10^6$~counts~s$^{-1}$~arcmin$^{-2}$ \citep[in \rosat\ units, see][]{snowden1997}. This figure is representative of the average background at low Galactic latitudes, where the \cite{bonamente2002} sample was selected, in order to avoid a region with a high Galactic column density that would prevent the observation of soft X--ray fluxes.
For the R34 \rosat\ band, we used a PSPC effective area of 50~cm$^2$ and approximated it with the 0.5-1~keV band, and we considered a typical diffuse background of 100-200~$\times 10^6$~counts~s$^{-1}$~arcmin$^{-2}$.
The R34 band soft-X--ray sky is considerably less variable than in the R12 band, and this background level is generally representative of most sight lines (see Fig.~\ref{fig:SxDetectability}).

The surface brightness of the WARM gas, especially in the $\nicefrac{1}{4}$ keV band, can be substantially lower than the typical diffuse background. For the more massive cluster (M4, right panel), the WARM gas surface brightness is comparable to the background in the inner cluster regions. Near the virial radius, the WARM gas is lower by at least a factor of 10 than the sky background. This means that accurate soft excess fluxes can only be measured using a careful estimate of the background, preferably using contemporaneous and in situ background measurements, as was the case for the \cite{bonamente2002} and \cite{bonamente2022c} observations. 
Moreover, the $\sim 0.5-1$~keV diffuse background is lower and more stable than in the $0.2-0.4$ keV band, as shown by the \rosat\ All--Sky Survey maps of  \cite{snowden1997}, for instance, and it is therefore more suitable for cases when local background is not available.

For the detection of the soft excess in the inner regions, the main advantage is provided by the inherently higher surface brightness of the WCGM, which is the dominant soft excess--producing phase (Fig.~\ref{fig:SxDetectability}). 
The soft excess in the inner regions is therefore less affected by the background subtraction \citep[see, e.g.,][]{bonamente2002, nevalainen2007}. Challenges for the detection in the inner regions are associated with the accurate determination (and therefore, removal) of the ICM gas because the soft excess is only expected to be a few percent above the surface brightness of the hot ICM (Fig.~\ref{fig:SxMassPlot}). 
In the inner cluster regions, the ICM temperature and abundance structure can be complex \citep[e.g.][]{rasia2014,mernier2017}, including for the presence of cool cores. This leads to systematic uncertainties in the detection of soft fluxes from warm subvirial gas.

Near the virial radius, the surface brightness of the soft excess becomes comparable to that in the hot ICM, and therefore, the details of the hot ICM are less important for the detection of the soft excess. At these radii, the main challenge is an accurate background subtraction because the soft excess is typically lower than the diffuse soft X--ray background. The ideal means for the detection of the soft excess near and beyond the virial radius are large--scale surveys such as the \rosat\ All--Sky Survey \citep[RASS, ][]{voges1999} or the \eRosita\ survey \citep[e.g.][]{liu2022} because of their contemporaneous and in situ background measurements. The analysis by \cite{bonamente2009} of the RASS data near Coma confirmed the detection of the soft excess from pointed observations \citep{bonamente2003}, and they tentatively extended the detection of the soft excess halo out to $\sim 2-3 \times r_{200}$. 
The soft excess emission is therefore detectable in the inner cluster regions and near the virial radius. This is consistent with available observations.

 To date, no measurements of the soft excess are available in the region beyond the virial radius.
As a proof of concept of the detectability of the soft excess in the $1-2 \times r_{200}$ region, we considered the 0.2-1 keV band (right panel of Fig.~\ref{fig:SxDetectability}). In this region, the WARM surface brightness depends on the mass of the host cluster, and it is approximately $\log S_X \sim -6.5$.
The emission of the most massive Coma--like clusters is higher, and we therefore assumed  $\log S_x = -6.5$ as a reference value
for a background of  $\log S_{x,b} = -5.5$ (combination of the R12 and R34 band backgrounds), that is, the WARM signal is 10\% of the average background level. 
For clusters at $z=0.1$ and with $r_{200}=1.5$~Mpc, it is immediately clear that a 100~ks observation with the \rosat\ PSPC in the 0.2-1 keV band yields $7.3\pm0.1 \times 10^3$~source counts against a background of $73\pm0.3 \times 10^3$~counts. When other sources with systematic errors associated with the background subtraction can be controlled, this exposure time is more than adequate to deliver a detection of the WHIM in emission from the circumcluster environments. Similar considerations can be applied to the \eRosita\ survey, for example, but with a somewhat higher background.~\footnote{It goes beyond the scope of this paper to provide a feasibility of WHIM detection with X--ray missions. However, \rosat\ has a $\geq 99$\% particle background rejection \citep{plucinsky1993}, which makes it uniquely suited for soft X--ray detections.}

These \tng\ data should not be extrapolated much beyond the virial radius in order to predict the expected count rates of the WARM and WHIM gas. The data we considered to make the radial profiles
only include particles within $5 \times r_{200}$, and no outer layers of the gas were included in the calculation of the projected surface brightness profiles \citep[see Paper~I, ][]{gouin2023}. Beyond the virial radius, the WHIM gas becomes preferentially distributed in filamentary structures, as shown by \cite{galarraga2020, galarraga2021, tuominen2021}. Therefore, it becomes more meaningful to seek the detection of this warm gas in specific filament regions and not in azimuthally averaged annuli, as we did here.

\section{Possible origins of the soft X--rays excess \label{sec:Origin}}

In Sect. \ref{sec:methods} we computed the thermodynamic property profile of TNG clusters (see Fig. \ref{fig:THERMO}) to predict the cluster X-ray emission and the soft-X-ray excess by separating the HOT and WARM gas phase (see Fig. \ref{fig:SxMassPlot}). These predictions match recent and previous observations from ROSAT, XMM, and eRosita well, as shown in Sect. \ref{sec:comparison} (see Figs. \ref{fig:SxPlot} and \ref{fig:ROSAT}). As suggested by \cite{gouin2023}, the clumpy WCGM gas is required in addition to the diffuse WHIM gas to explain the soft excess in observations. In this section, we aim to understand the origin of the soft X--ray excess better as a function of the cluster distance.

\begin{figure}[h]
    \includegraphics[width=3.5in]{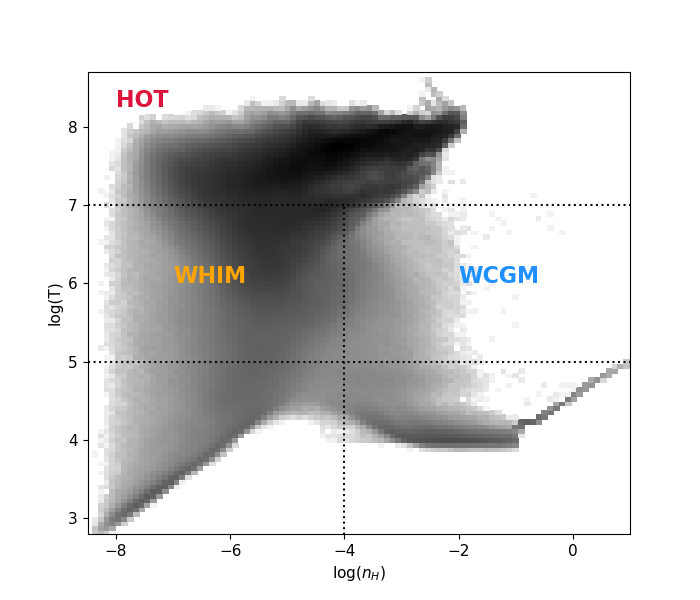}
    \caption{Temperature-density diagram for all gas cells around a representative \tng\ cluster up to 5 $r_{200}$, with the three gas phases of interest, HOT, WHIM, and WCGM.}
    \label{fig:Illustration}
\end{figure}
\begin{figure*}[h]
    \centering
    \includegraphics[width=7in]{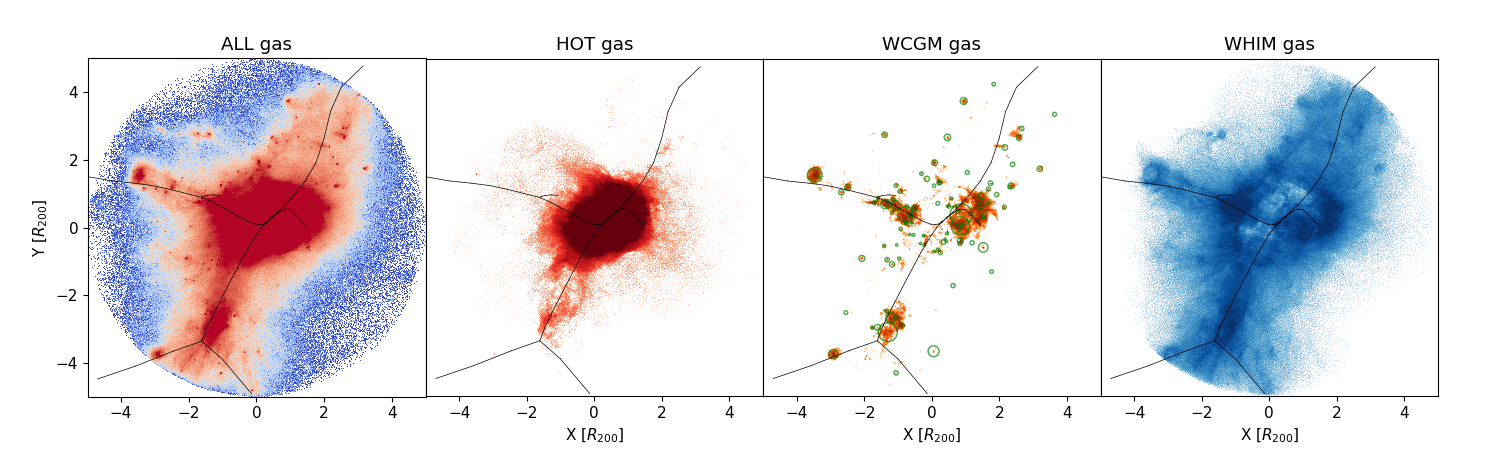}
    \caption{Gas distribution as a function of the different phases as defined in Table~\ref{tab:gas} around a representative \tng\ cluster. 
    We overplot the cosmic filaments detected by T-REX algorithm as dark lines. We overplot the galaxies (with a stellar mass $M_{*}>10^{9} M_{\odot}/h$) as green circles with radii proportional to the galaxy mass. }
    \label{fig:Illustration2}
\end{figure*}

\subsection{Visualizing WCGM and WHIM gas}

To understand the origin of the soft X--ray emission better, we illustrate in Fig.~\ref{fig:Illustration} and Fig.~\ref{fig:Illustration2} the three main gas phases (HOT, WCGM, and WHIM gas) and their distribution around a simulated cluster. 
As expected, the clumps of WARM gas are in the form of WCGM gas because this phase is by definition high-density warm gas. The other phase that contributes to the WARM gas, the WHIM (rightmost panel) shows a rather smooth spatial distribution. Interestingly, this WCGM gas apparently strongly populates the subhalos of galaxies inside and around clusters, as illustrated by the green circles, which show galaxies with a stellar mass $M_{*}>10^{9} M_{\odot}/h$. The radius of the green circles is equal to twice the radius containing half of the total mass of each galaxy. These galaxies correlate very well with the spatial distribution of WCGM gas, as expected from the lower-density threshold that defines this gas phase \cite{martizzi2019}. In agreement with \cite{angelinelli2021}, the gas clumps tend to trace filamentary patterns around clusters. The cosmic web skeleton, shown as black lines, was computed by using the algorithm called TREX filament finder \citep{Bonnaire2020} on 3D galaxy distribution, similarly to \cite{gouin2022}. The right panel of Fig.~\ref{fig:Illustration} qualitatively confirms that the WHIM gas traces the filaments that are connected to clusters.

\subsection{Definition and computation of the gas fraction}

\subsubsection*{The gas-phase fraction}

To explore the effect of the gas-phase abundance on the soft X--ray excess, we computed its mass fraction as
\begin{equation}
    f_{i} (r)= \frac{M_{i} (r)}{M_{\mathrm{all \ gas}} (r)} \,,
\end{equation}
with $i$ the gas phase (WCGM or WHIM), and $r$ the clustercentric radius enclosing the total gas mass ($M_{\mathrm{all \ gas}}$) and the mass of 
the $i$th gas-phase ($M_{i})$.
We chose a radial range from $R=0$ to $5 \ r_{200}$ in intervals of $1 \times \ r_{200}$ to accurately probe the radial evolution.

\subsubsection*{The gas fraction in filaments}
 
To explore whether the soft X--ray emission is the result of either WCGM gas surrounding galaxies and/or WHIM gas inside filaments, we selected the gas inside filaments and/or inside subhalos.
We considered gas located in filaments to be gas cells enclosed within 2 Mpc/h around the filament spine \citep[which is a conservative value for the typical radial thickness of filaments based on their density profiles ][]{galarraga2020, Wang2024}. We excluded gas cells from galaxies, which are defined as those inside twice the half-mass radius of each galaxy with a stellar mass $M_*>10^9 \ M_{\odot}/h$. The cosmic filaments were detected with the T-REX algorithm in galaxies with a stellar mass $M_*>10^9 \ M_{\odot}/h$ inside spheres with a radius of $10 \times r_{200}$ centered on each cluster \citep[similar to][]{gouin2021}. 

We quantified the gas in filaments via
\begin{equation}
    f^{\rm fil}_{\rm gas} (r)= \frac{M^{\in \mathrm{filaments}}_{\mathrm{all \ gas}}(r)}{M_{\mathrm{all \ gas}}(r)} \,.
\end{equation}
For each cluster, we removed the contribution of gas that was  included in galaxies that were located in filaments to investigate the effect of galaxies and filaments on the soft X--ray excess separately.

\subsubsection*{The gas fraction in substructures}

Similarly, we computed the fraction of gas that surrounded massive galaxies ($M_*>10^9 \ M_{\odot}/h$) as
\begin{equation}
    f^{\rm gal}_{gas} (r)= \frac{M^{\in \mathrm{galaxies}}_{\mathrm{all \ gas}}}{M_{\mathrm{all \ gas}}} \,.
\end{equation}
 For each cluster, we did not consider the main central galaxy to exclude the contribution of the cluster itself. Again, we considered the gas cells in galaxies to be the gas inside twice the half-mass radius of each galaxy.
 
\subsubsection*{The subhalo mass fraction}

Finally, the subhalo mass fraction was estimated from
\begin{equation}
    \label{eq:fsub}
    f_{\rm sub} (r)= \frac{ \sum_{\rm sub \in r} M_{\rm sub}}{M_{\rm main}} \,,
\end{equation}
where  $M_{\rm sub}$ is the total mass of a given subhalo inside the radial annulus $r$, and $M_{\rm main}$ is the total mass of the main subhalo for each cluster.  Only for the computation of $f_{\rm sub}$ did we consider all the subhalos extracted by the {\it Subfind} algorithm (without any mass selection) to fully capture the mass contained in clumps. It therefore reflects the relative clumpiness of the matter distribution.

\subsection{Correlation of the soft X--ray excess with the gas properties}
\label{sec:correlation}
\begin{figure}[h]
    \centering
    \includegraphics[width=3.45in]{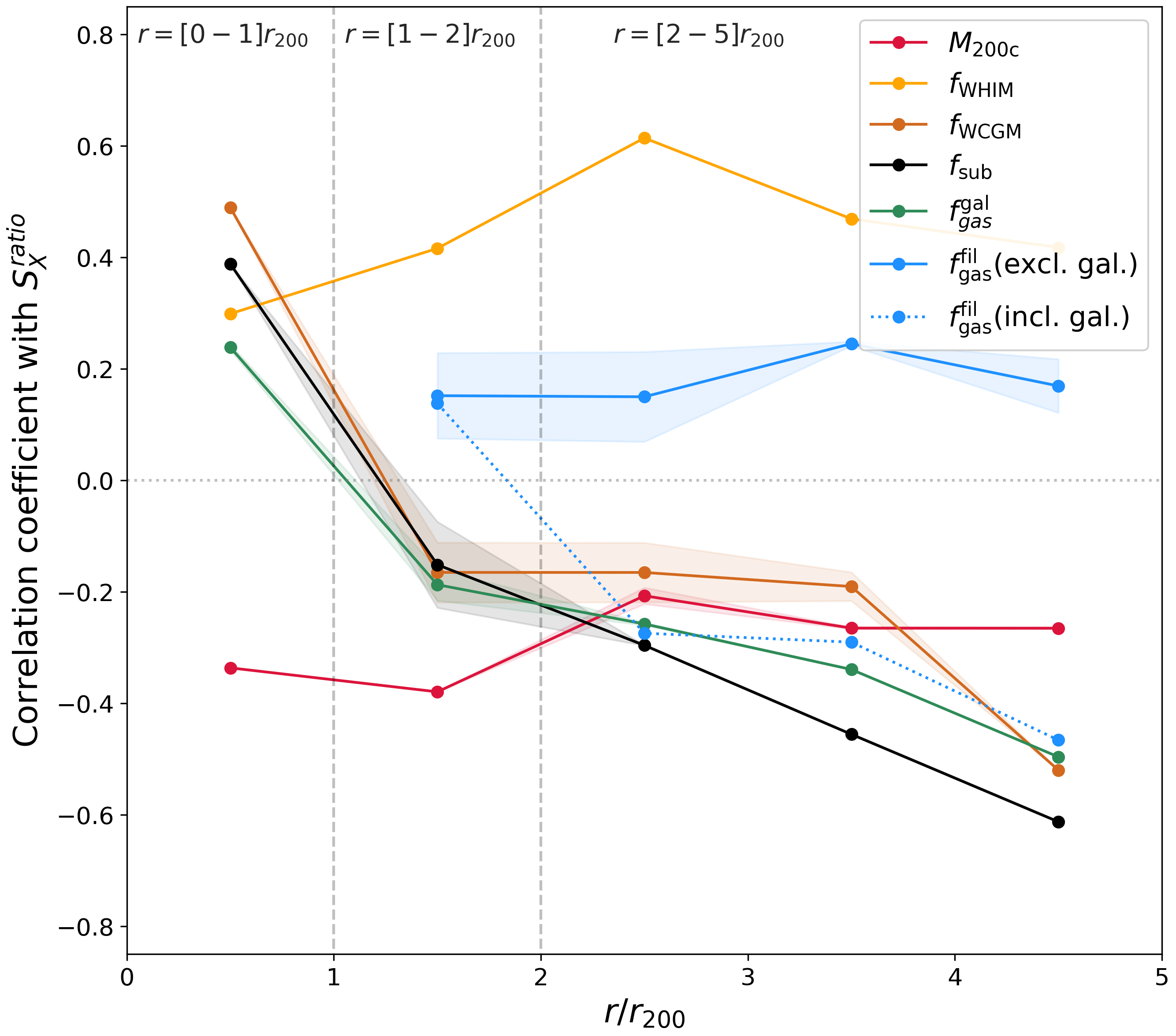}
    \caption{ Radial evolution of the Spearman correlation coefficient between the soft X--ray excess ($S^{ratio}_X$) and gas properties as a function of cluster distance $R$. The $p$ values are represented by the error bars. 
     We present the direct correlation with cluster masses (red line), but the other properties were computed by using the partial correlation to first remove any mass dependence.}
    \label{fig:properties_R}
\end{figure}

\begin{figure}[h]
    \centering
    \includegraphics[width=3.45in]{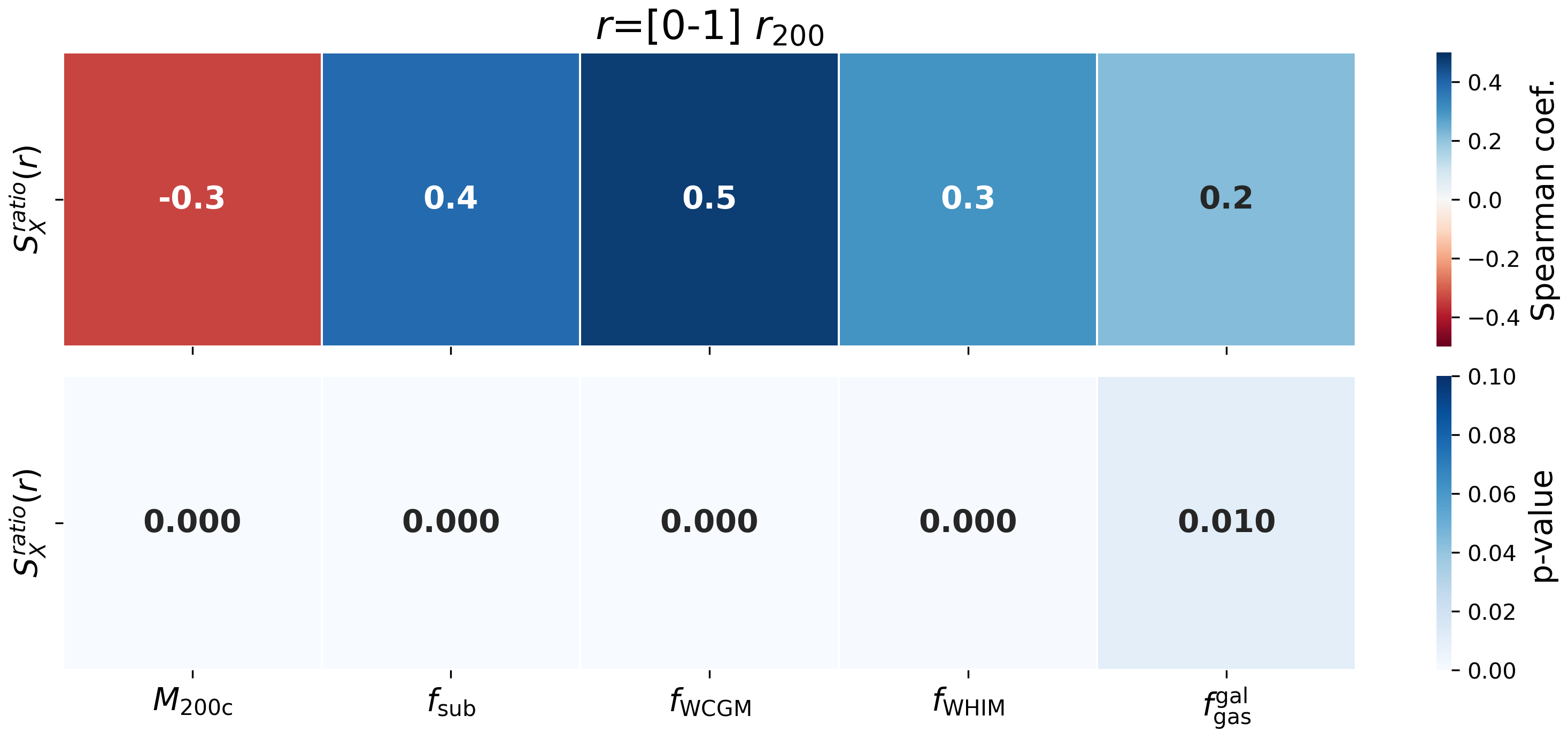}\
    \includegraphics[width=3.45in]{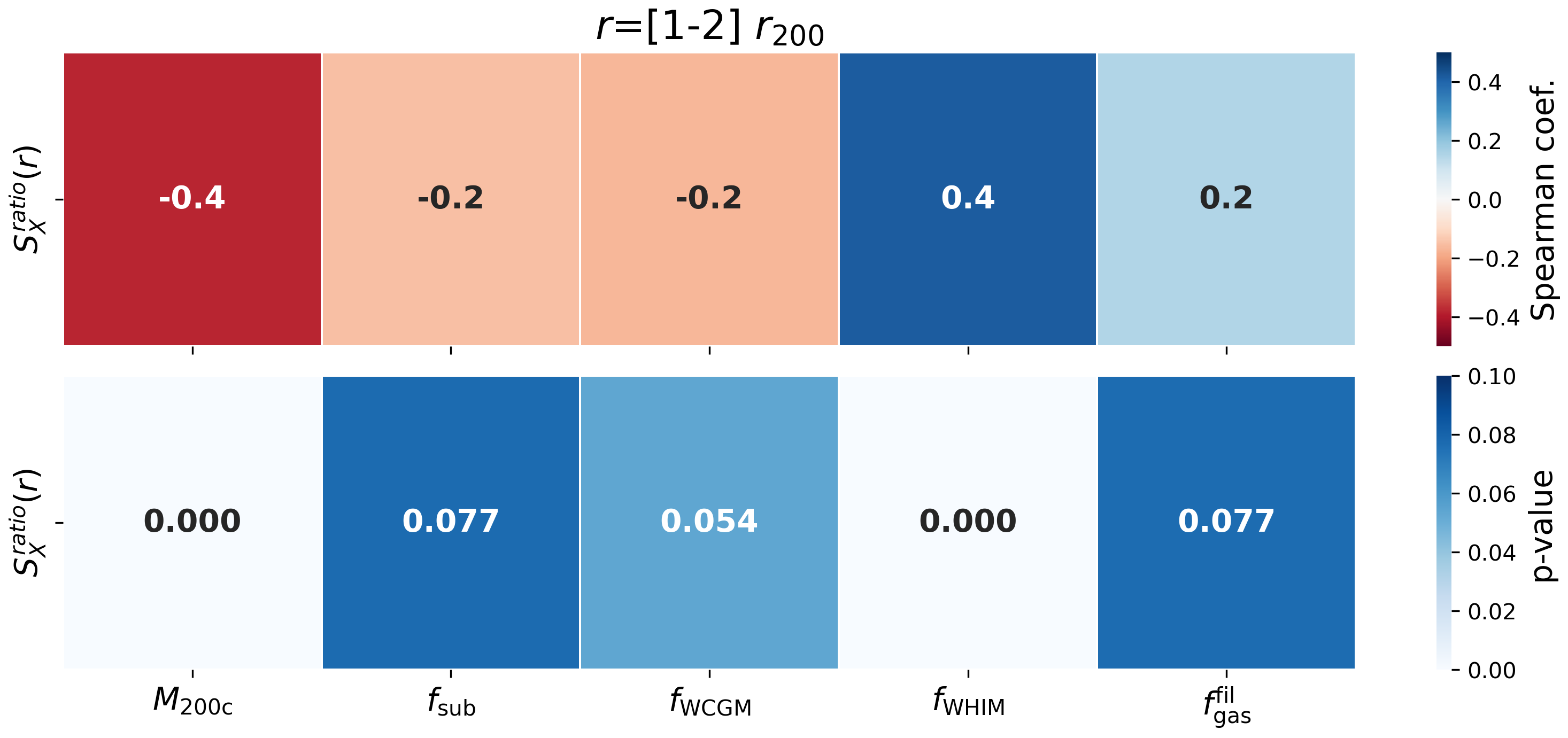}\
    \includegraphics[width=3.45in]{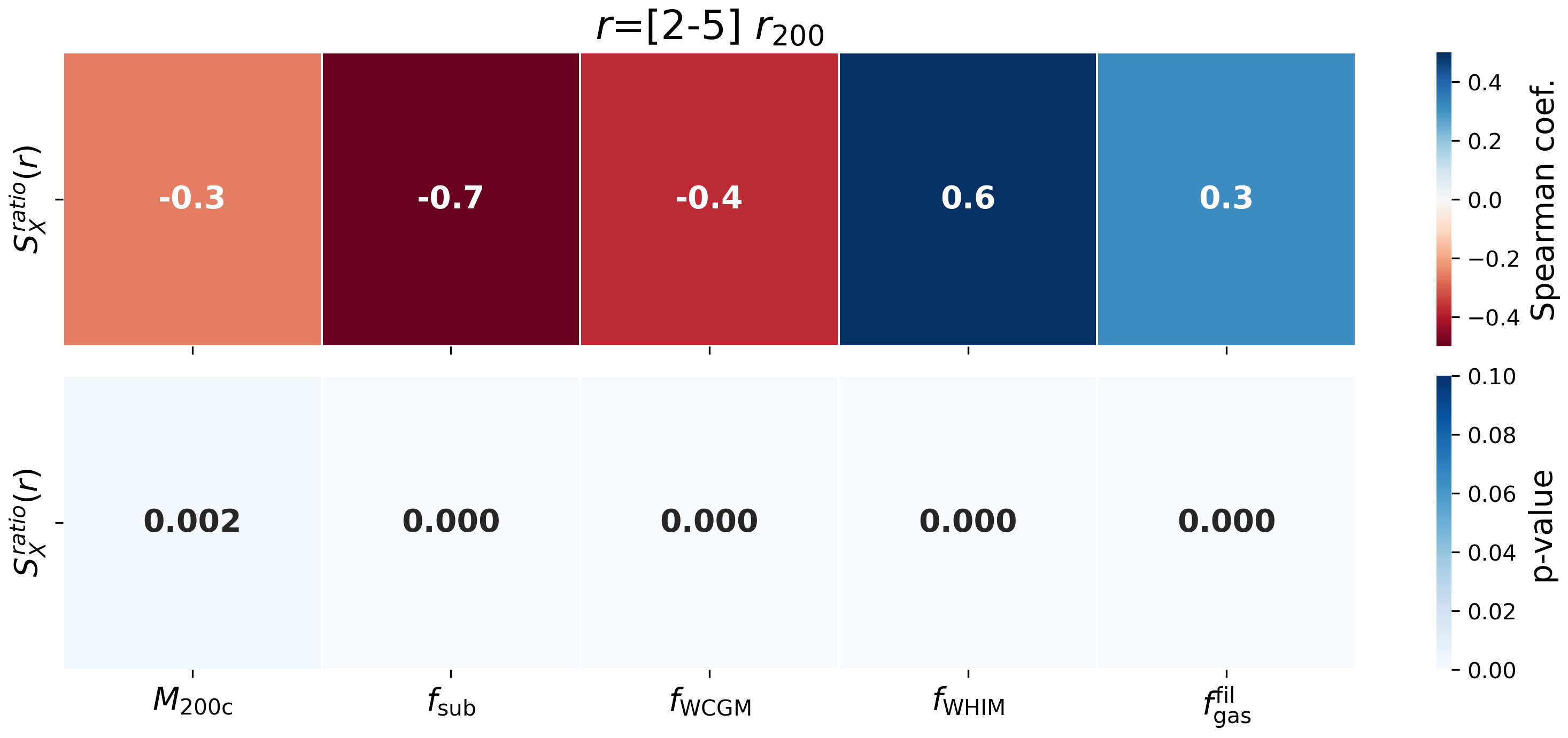}\
    \caption{Spearman correlation coefficient between the soft excess and the cluster gas properties and its associated $p$ values, integrated over three radial ranges: inside clusters $0>R[r_{200}]>1$, around clusters $1>R[r_{200}]>2$, and outside clusters $2>R[r_{200}]>5$. As in Fig~\ref{fig:properties_R}, except for the cluster mass ($M_{200}$), we show the partial Spearman correlation to remove any mass dependence.}
    \label{fig:properties_R123}
\end{figure}

For each cluster, we integrated the soft X--ray excess, which is 
defined as the ratio of the WARM to the HOT gas X--ray emission and is hereafter 
referred to as $\log S^{ratio}_{X}$ , inside different radial annuli centered on the clusters with radial ranges of $[0-1],[1-2],[2-3],[3-4]$, and $[4-5]\ r_{200}$. We analyzed the correlation between the integrated soft X--ray excess and the gas properties by computing the Spearman correlation coefficient and its associated $p$ value. The integrated soft X--ray excess presented Fig~\ref{fig:properties_R} and Fig~\ref{fig:properties_R123} wa calculated for the energy band $0.2$-$1.0$ keV. 

In Fig.~\ref{fig:properties_R} we show the correlations between the soft X--ray excess and the gas properties as a function of cluster distance. 
We observe a smooth radial evolution in the soft excess influences, with a gradual transition. As detailed in Fig.~\ref{fig:properties_R123}, this can be divided into three radial regimes: inside clusters at $r<r_{200}$ (top panel), near the virial radius at $1<r/ r_{200}<2$ (middle panel), and outside clusters at $r >2 \times r_{200}$ (bottom panel). 
The soft X--ray excess in the two figures is anticorrelated with the cluster mass in all radial ranges. As anticipated from Fig.~\ref{fig:SxMassPlot}, clusters with a lower mass exhibit a greater soft excess. Except for the cluster mass, all correlations are therefore the partial correlation coefficient, which is defined as the degree of association between the soft excess and the gas properties after the effect of the primary variable is removed, that is, the dependence on mass. 
Beyond this mass-driven effect, we interpret the correlations between the soft X--ray excess and gas properties as listed below.
\begin{itemize}
    \item[(i)] Inside clusters ($r<r_{200}$), the soft X--ray excess correlates strongly with the WCGM gas fraction, the fraction of substructures, and the fraction of gas inside satellite galaxies. This suggests that the soft X--ray excess is mainly produced by substructures that host WCGM gas (with a correlation factor of 0.5 in the top panel of Fig~\ref{fig:properties_R123}). The WHIM gas inside clusters also affects the soft X--ray excess, but with a lower correlation factor of $0.3$. 
    \cite{Marini2025} recently discovered that luminous soft X--ray galaxy groups are more recently formed than their fainter counterparts with similar masses. Consequently, we speculate that the soft excess inside clusters, which correlates with the substructure fraction, is a proxy for a different dynamical state that is induced by a different mass-assembly history. We recall that \cite{Marini2025} explored the soft X-ray luminosity, and not specifically the soft X-ray excess.
    \item[(ii)] Around clusters ($1<r/r_{200}<2$), the role of the WHIM gas becomes more significant, with a correlation factor of $0.4$. 
    The correlation with other properties has high $p$ values, however, which makes it difficult to distinguish the effect of WCGM gas clumps and WHIM filaments on the soft excess in this particular regions, as shown in the middle panel of Fig~\ref{fig:properties_R123}.
    The reason might be that these cluster boundaries are transitioning regions, in which WARM gas is in-falling, out-flowing, and shocking during the accretion process \citep{Rost2021}.
    \item[(iii)] Outside clusters ($2<r/r_{200}<5$), the soft excess is mostly induced by WHIM gas inside cosmic filaments. We found that the soft X--ray excess tends to correlate with the fraction of gas in filaments (only when the gas in galaxies is excluded), and it correlates strongly with the WHIM fraction (blue and yellow points in Fig.~\ref{fig:properties_R}). 
    The strong anticorrelation with the WCGM and other substructure fractions also suggests that the more diffuse the gas, the higher the soft X--ray excess in this region.

\end{itemize}

\begin{figure*}[h]
    \centering
    \includegraphics[width=6in]{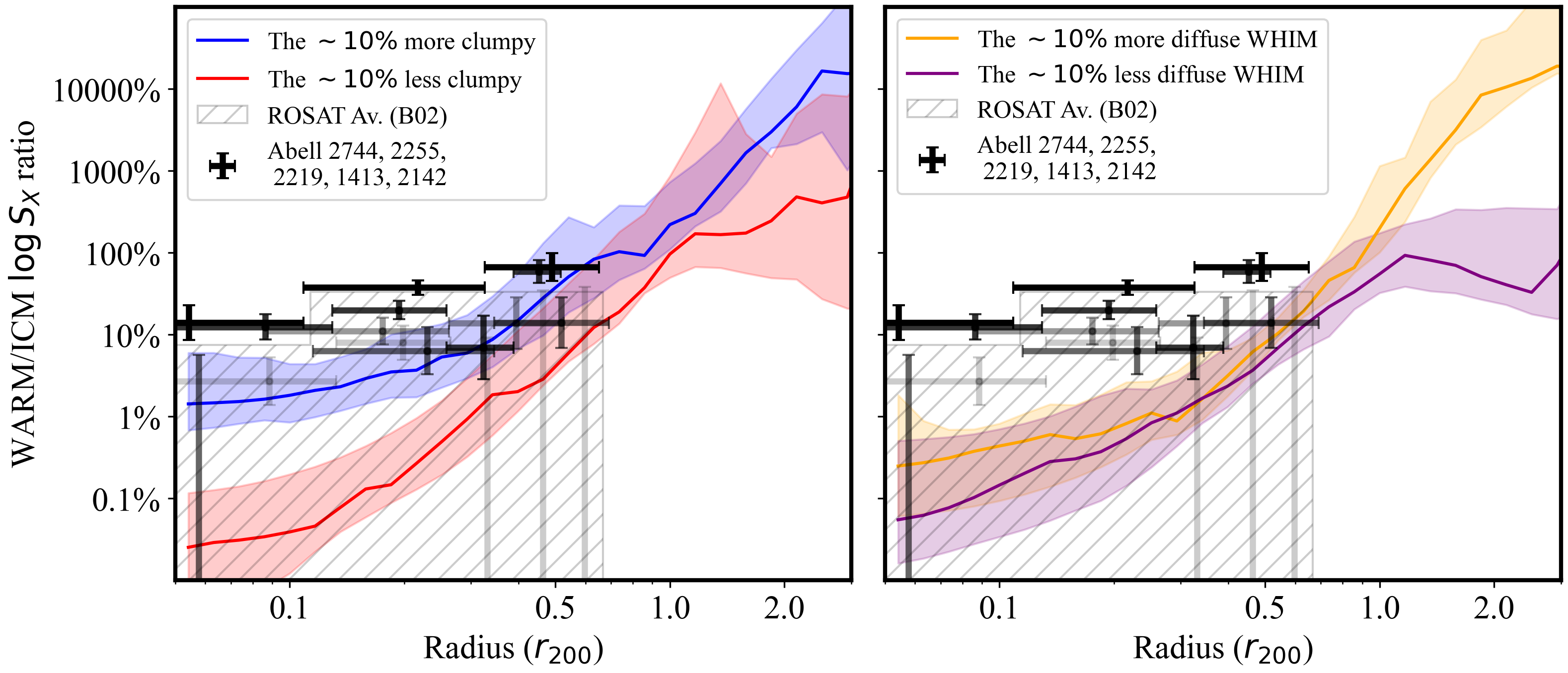}
    \caption{Median soft excess of four cluster populations with different gas-phase and substructure fractions (details in Sec.~\ref{sec:correlation}). The error bars represents the percentiles from 20\% to 80\%.}
     \label{fig:relax_Sx}
 \end{figure*}

To further test whether the soft excess results from a substructure inside the clusters and
from the diffuse WHIM outside, we performed the following two selections.
First, we defined less clumpy systems as those in the lowest 25\% of the substructure fraction and WCGM fraction inside $r_{200}$, indicating spatially clumpy clusters. We obtained a selection of 14 clusters. Conversely, the more clumpy systems were selected as those in the highest 25\% of both parameters, representing clumpy clusters with a larger fraction of dense clumps. This resulted in a selection of 15 clusters. 
The left panel of Fig.~\ref{fig:relax_Sx} confirms our finding that the soft X-ray excess in unrelaxed clusters is significantly higher inside $0.5 \ r_{200}$ than in relaxed objects. 
Second, we considered systems that are strongly populated by the diffuse WHIM outside the virial radius (between 1 and 5 $\times r_{200}$) as those in the lowest 25\% of the substructure fraction and the highest 25\% of the WHIM fraction. This  resulted a selection of 21 clusters with a more diffuse WHIM. Conversely, systems with a less diffuse WHIM in their outskirts were selected as those in the highest 25\% of the substructure fraction and the lowest 25\% of the WHIM fraction outside the virial radius, representing less diffuse and more clumpy cluster environments. This  resulted a selection of 12 clusters.
The right panel of Fig.~\ref{fig:relax_Sx} confirms that the soft excess in cluster outskirts that are strongly populated by WHIM gas is significantly higher at radii $>1 \times r_{200}$ than for those with clumpy environments.

\section{Discussion \label{sec:Disc}}

\subsection{Interpretations of the soft excess inside clusters}

The soft excess detected inside clusters can be explained by the thermal emission of subvirial WARM gas. Inside the virial radius, the WARM gas is substantially more clumpy that the hot ICM (see Fig. \ref{fig:THERMO}), with a significant mass fraction (from 0.1\% to more than 10\%; see Fig. \ref{fig:MASS}) producing the inner soft excess at a level of 0.1 to more than 10\% of the hot ICM emission (see Fig. \ref{fig:SxMassPlot}). Its large scatter is consistent with the detection of only $\sim 40$\% of clusters with a soft excess with \rosat\ (see Fig.~\ref{fig:ROSAT}).

Exploring the origin of the inner soft X--ray excess, we found that the large observed scatter might be explained be the large variety of dynamical cluster states (see Fig.~\ref{fig:relax_Sx}). We found that the soft excess is stronger for the more clumpy clusters. This trend is significant, and the excess decreases from 1–10\% for the more clumpy objects to less than 0.1\% for the most homogeneous ones (see Fig.~\ref{fig:relax_Sx}). 
Our results suggests that the soft X--ray excess is mainly produced by substructures that host WCGM gas (see Fig.~\ref{fig:Illustration}). This interpretation of a soft X--ray boost inside clusters also agrees with previous studies that found that inhomogeneities boost the estimates of the X-ray emissivity \citep[see e.g.][]{Planelles2017,Ghirardini2018}.

Systems that experienced a recent merger (i.e., that are still below the virialization timescale) are expected to be more clumpy, with residual BCGs and satellites originating from the two merging halos. These substructures are expected to be surrounded by WARM gas in the form of WCGM (as illustrated in Fig~\ref{fig:Illustration2}). In these recently formed systems, the ICM is expected to be not yet fully virialized, but rather in multiphase and inhomogeneous. A contrast might therefore occur between the hard X-ray emission from the virialized hot ICM and the soft emission produced by the inhomogeneous WARM gas that is still being mixed and incorporated into the hot component.
Recently, \cite{Marini2024} found that X-ray bright objects in their cores (i.e., within $r_{500}$) exhibit rapid mass accretion during the later stages of their evolution in the Magneticum simulation \citep{Magneticum}.
Additionally, \cite{gouin2022} demonstrated a strong correlation between the formation time, accretion rate, and degree of relaxation of a cluster with the asymmetries in its gas distribution.
Based on these studies, we infer that a high excess of soft X-rays is related to an unrelaxed cluster as a consequence of a late-stage mass-assembly history. In contrast, relaxed clusters that formed early are expected to preferentially have a lower soft excess.

 Even though we tended to reproduce most of the soft excess in the central regions as thermal emission from warm gas clumps, we remain limited by the resolution of the simulation ($\sim$ kiloparsec). In addition to the contribution from unresolved clumps, we also did not account here for nonthermal processes, such as turbulence and shocks, that are injected into the ICM  \citep{VallesPerez2021,Lebeau2025}. 

\subsection{Interpretations of the soft excess around clusters}

Near the virial radius, the WARM gas density dominates the hot ICM. The exact value of this transition from ICM to WARM  depends on the cluster mass (see Fig.~\ref{fig:MASS}) in such a way that massive clusters have an ICM phase that is more spatially extended than low--mass clusters. This finding suggests that the soft excess might preferentially occur at larger radii in more massive clusters, as was indeed observed in the case of Coma (see Fig.~\ref{fig:ROSAT}, \citealt{bonamente2003}).
The Coma cluster is the only case to date for which a strong soft excess emission was detected up to the virial radius. The study by \cite{bonamente2002} of a large sample of \rosat\ clusters did not attempt to measure the soft excess beyond this radius, where the data were dominated by background (see Fig.~\ref{fig:SxDetectability}), and therefore, there is no statistically significant cluster sample so far for which these predictions have been tested. Our predictions of the soft X--ray excess reproduce the Coma observations near the virial radius very well by considering WCGM clumps and diffuse WHIM gas. This result is consistent with the finding by \cite{churazov2023} that the Coma soft excess cannot be reproduced by assuming WHIM gas alone.

Further insights into the origin of the soft X-ray excess in this region remain challenging, as evidenced by the fact that only the WHIM fraction shows a strong correlation with the excess (see Fig.~\ref{fig:properties_R123}). This is due to complicated physical processes in the cluster peripheries that are induced by diffuse material that collapses toward filaments and gas-shock processes triggered by substructures.
\cite{Rost2021} have studied the complex velocity field of gas in this regions in detail and showed that gas preferentially enters the cluster as part of a filament and leaves the cluster outside filaments \citep[see also][]{Rost2024}.
Additionally, the main gas-accretion shock in the clusters is expected in this radial regions, for instance, near  $1.5 \times r_{200}$, according to \cite{Arthur2019}. Similarly, \cite{Vurm2023} have also found the formation of a gas shock where filaments were connected the hot atmosphere of clusters in the C-EAGLE simulation. \cite{Power2020} demonstrated that baryonic physics models of simulations converge on the time and location of shocks, but differ in the shock strength. Moreover, \cite{Lebeau2024} discussed the distinction between gas and the dark matter splashback radius and argued that the detection of the splashback radius in pressure profiles might be more related to an accretion shock.
In this line, \cite{churazov2023b} have studied the complex geometry of the accretion shock in this region around the Coma cluster.
This transition from HOT-to-WARM gas contains a complex mixture of infalling and outflowing gas, including potential backsplash gas \citep{gouin2022,Kotecha2022}. This complicates the study of this peripheral region.

\subsection{Interpretations of the soft excess outside clusters}

It is important to point out that the meaning of the soft X--ray excess outside clusters ($>2 \ r_{200}$) 
is different from its usual definition as an excess of soft-X emission above the hot ICM inside a given cluster. In this circumcluster medium, the excess  now refers to the relative importance 
between soft X--ray emission from warm and that of hot gas induced by any structures that are neighbor to the central cluster, and not just the central cluster itself. An investigation
of this region will be the subject of a follow--up paper, which  will focus on the X-rays emission by distinguishing the structures (clusters, groups, filaments, and clumps) instead of the current X-ray ratio of the gas phases. This study goes beyond the scopes of this paper.

In general, we predict that clusters of all masses feature a total WARM mass of $\sim 2 \times 10^{13}$~M$_{\odot}$ within $2 \times r_{200}$, which corresponds to approximately 10\% of the ICM mass for the more massive clusters and to $\sim 50$\% for the low-mass clusters (see Fig.~\ref{fig:MASS}). 
We showed that at $5\times r_{200}$, the WARM halo mass is comparable to that of the hot ICM, in particular, for the cluster mass range $M_1$ and $M_2$ with $\log M <14.5$.
This indicates that the circumcluster warm gas is a significant reservoir of baryons, of which the soft excess emission near the virial radius represents only a very small part. This outer region from 2 to 5 $r_{200}$ is dominated by WHIM gas that is mostly located inside cosmic filaments \citep{galarraga2021, tuominen2021}. 
At this scale, our findings indicate that the soft excess is correlated with the WHIM fraction and gas abundance in filaments (see Fig.~\ref{fig:properties_R123}). This shows that lower fraction of substructures leads to a higher soft excess (see Fig.~\ref{fig:relax_Sx}).

\subsection{Detectability of warm gas in cluster peripheries}

   Only a few detections of individual WHIM filaments in emission were made so far, starting from  filaments toward superclusters \citep{kull1999,zappacosta2005}, so-called spider legs around Abell~2744 \citep{eckert2015,Gallo24}, to WHIM bridges between cluster pairs \citep{sakelliou2004,werner2008}.
   More recently, the detection of a soft-X--ray filament between Abell~3391 and Abell~3395 with \eRosita\ was reported by \cite{reiprich2021} and was also studied by \cite{alvarez2018}, and a filament between Abell 2029 and Abell 2033 was also reported \citep{mirakhor2022} that was previously identified from \rosat\ \citep{walker2012a}. 
   Recently, \cite{Dietl2024} identified an X-ray excess beyond thrice the virial radius of Abell 3667, which is indicative of an intercluster filament. Recent research focused on stacking intercluster gas filaments to enhance the signal-to-noise ratio for detecting warm gas \citep{Tanimura2022,Zhang2024_erosita}. 
   Another possibility is the detection of the WARM gas via their  emission lines. 
   \cite{Zhang2024} recently used the 30 most massive and relaxed \tng\ clusters to show that they enable the use of emission lines for measuring the thermodynamic, chemical, and kinematic properties of the gas up to $r_{200}$ and beyond with future missions, in particular, with the Line Emission Mapper (LEM) \citep[see also][]{Tuominen2023}. For further exploration, \cite{Zhao2025} recently provided forecasts for WHIM detections outside clusters with the  proposed mission called Hot Universe Baryon Surveyor.
 Our results indicate that the soft X--ray emission from WARM gas in the cluster environment might be detectable toward the region immediately outside of the virial radius, as shown in Sec.~\ref{sec:detectability}.
 At these radii, the gas is mostly in the form of diffuse WHIM filaments, which are known to have a higher density toward galaxy clusters \citep{Li2025}. We argue that the region outside of the virial radius of massive clusters is ideal for detecting WHIM filaments in emission. The recent detection of soft X--ray emission near clusters \citep[e.g.][]{reiprich2021, mirakhor2022, Dietl2024} supports this scenario and indicates that the circumcluster environment is the region in which we should search for X--ray emission of filaments. 

\section{Conclusions \label{sec:conclus}}

 We investigated the origin of the soft X--ray emission in and near \tng\ galaxy clusters. 
 In general, our results suggest that the WARM gas phase causes an excess of soft X--ray emission above the contribution from the virialized HOT phase, as was observed by a variety of X--ray missions \citep[e.g.,][]{lieu1996a, lieu1996b, bonamente2002,bonamente2003}. In detail, we distinguished three different radial regimes with different gas structures and properties that cause this soft excess.
 In the inner cluster regions, this soft excess is primarily expected to be due to the high--density and high--clumpiness WCGM phase. Toward the virial radius, the soft excess is given by a combination of two distinct WARM gas phases, that is, the high--density WCGM gas and the low--density WHIM gas. The soft X--ray predictions for the most massive clusters match the observations of the Coma cluster very well, which is the only cluster to date for which soft excess emission to the virial radius was reported \citep{bonamente2003, bonamente2009, bonamente2022c}. 
 Beyond the virial radius, the diffuse WHIM phase becomes the dominant driver of the soft excess in cluster environments. This is consistent with the established picture of galaxy clusters being the nodes at which WHIM filaments converge \citep[e.g.][]{cen1998, dave2001}. The soft X--ray emission from WHIM filaments that project toward the region immediately outside the virial cluster radius  (i.e., within a few $\times r_{200}$) is a promising avenue for the detection of WHIM filaments in emission. 
 This radial picture is the natural way to interpret the soft excess as due to WCGM in the inner regions, a combination of WCGM and WHIM near the virial radius \citep[e.g., as in Coma][]{bonamente2022c}, and WHIM beyond the virial radius \citep{reiprich2021, mirakhor2022, Dietl2024}.
 Our analysis offers additional clues about the physical conditions that favor soft excess. In particular, clusters with a larger substructure fraction are expected to have a higher central soft excess, and clusters surrounded by a larger fraction of diffuse WHIM naturally show a high excess beyond the virial radius.  While additional details remain to be investigated, these findings provide a relatively simple first--order picture of the origin of the soft X--ray excess in circumcluster environments.
 
\begin{acknowledgements}
The authors thank an anonymous referee for their useful comments and suggestions. 
CG acknowledge funding from the French Agence Nationale de la Recherche for the project WEAVEQSO-JPAS (ANR-22-CE31-0026).
We thank the IllustrisTNG collaboration for providing free access to the data used in this work. 
This work was supported in part by NASA's Astrophysics Data Analysis Program (ADAP) grant {'Closing the Gap on the Missing Baryons at Low Redshift with multiwavelength observations of the Warm--Hot Intergalactic
Medium'} awarded to the University of Alabama in Huntsville.
\end{acknowledgements}

\bibliography{max}

\begin{thebibliography}{96}
\expandafter\ifx\csname natexlab\endcsname\relax\def\natexlab#1{#1}\fi

\bibitem[{{Allen} {et~al.}(2004){Allen}, {Schmidt}, {Ebeling}, {Fabian}, \& {van Speybroeck}}]{allen2004}
{Allen}, S.~W., {Schmidt}, R.~W., {Ebeling}, H., {Fabian}, A.~C., \& {van Speybroeck}, L. 2004, \mnras, 353, 457

\bibitem[{{Alvarez} {et~al.}(2018){Alvarez}, {Randall}, {Bourdin}, {Jones}, \& {Holley-Bockelmann}}]{alvarez2018}
{Alvarez}, G.~E., {Randall}, S.~W., {Bourdin}, H., {Jones}, C., \& {Holley-Bockelmann}, K. 2018, \apj, 858, 44

\bibitem[{{Angelinelli} {et~al.}(2021){Angelinelli}, {Ettori}, {Vazza}, \& {Jones}}]{angelinelli2021}
{Angelinelli}, M., {Ettori}, S., {Vazza}, F., \& {Jones}, T.~W. 2021, \aap, 653, A171

\bibitem[{{Arnaud} {et~al.}(2002){Arnaud}, {Aghanim}, \& {Neumann}}]{arnaud2002}
{Arnaud}, M., {Aghanim}, N., \& {Neumann}, D.~M. 2002, \aap, 389, 1

\bibitem[{{Arthur} {et~al.}(2019){Arthur}, {Pearce}, {Gray}, {Knebe}, {Cui}, {Elahi}, {Power}, {Yepes}, {Arth}, {De Petris}, {Dolag}, {Garratt-Smithson}, {Old}, {Rasia}, \& {Stevens}}]{Arthur2019}
{Arthur}, J., {Pearce}, F.~R., {Gray}, M.~E., {et~al.} 2019, \mnras, 484, 3968

\bibitem[{{Bahar} {et~al.}(2024){Bahar}, {Bulbul}, {Ghirardini}, {Sanders}, {Zhang}, {Liu}, {Clerc}, {Artis}, {Balzer}, {Biffi}, {Bose}, {Comparat}, {Dolag}, {Garrel}, {Hadzhiyska}, {Hern{\'a}ndez-Aguayo}, {Hernquist}, {Kluge}, {Krippendorf}, {Merloni}, {Nandra}, {Pakmor}, {Popesso}, {Ramos-Ceja}, {Seppi}, {Springel}, {Weller}, \& {Zelmer}}]{Bahar2024}
{Bahar}, Y.~E., {Bulbul}, E., {Ghirardini}, V., {et~al.} 2024, \aap, 691, A188

\bibitem[{{Barnes} {et~al.}(2017){Barnes}, {Kay}, {Bah{\'e}}, {Dalla Vecchia}, {McCarthy}, {Schaye}, {Bower}, {Jenkins}, {Thomas}, {Schaller}, {Crain}, {Theuns}, \& {White}}]{Barnes2017}
{Barnes}, D.~J., {Kay}, S.~T., {Bah{\'e}}, Y.~M., {et~al.} 2017, \mnras, 471, 1088

\bibitem[{{Bonamente} {et~al.}(2003){Bonamente}, {Joy}, \& {Lieu}}]{bonamente2003}
{Bonamente}, M., {Joy}, M.~K., \& {Lieu}, R. 2003, \apj, 585, 722

\bibitem[{{Bonamente} {et~al.}(2009){Bonamente}, {Lieu}, \& {Bulbul}}]{bonamente2009}
{Bonamente}, M., {Lieu}, R., \& {Bulbul}, E. 2009, \apj, 696, 1886

\bibitem[{{Bonamente} {et~al.}(2002){Bonamente}, {Lieu}, {Joy}, \& {Nevalainen}}]{bonamente2002}
{Bonamente}, M., {Lieu}, R., {Joy}, M.~K., \& {Nevalainen}, J.~H. 2002, \apj, 576, 688

\bibitem[{{Bonamente} {et~al.}(2001{\natexlab{a}}){Bonamente}, {Lieu}, \& {Mittaz}}]{bonamente2001}
{Bonamente}, M., {Lieu}, R., \& {Mittaz}, J.~P.~D. 2001{\natexlab{a}}, \apjl, 561, L63

\bibitem[{{Bonamente} {et~al.}(2001{\natexlab{b}}){Bonamente}, {Lieu}, \& {Mittaz}}]{bonamente2001c}
{Bonamente}, M., {Lieu}, R., \& {Mittaz}, J. P.~D. 2001{\natexlab{b}}, \apjl, 547, L7

\bibitem[{{Bonamente} {et~al.}(2005){Bonamente}, {Lieu}, {Mittaz}, {Kaastra}, \& {Nevalainen}}]{bonamente2005}
{Bonamente}, M., {Lieu}, R., {Mittaz}, J.~P.~D., {Kaastra}, J.~S., \& {Nevalainen}, J. 2005, \apj, 629, 192

\bibitem[{{Bonamente} {et~al.}(2001{\natexlab{c}}){Bonamente}, {Lieu}, {Nevalainen}, \& {Kaastra}}]{bonamente2001b}
{Bonamente}, M., {Lieu}, R., {Nevalainen}, J., \& {Kaastra}, J.~S. 2001{\natexlab{c}}, \apjl, 552, L7

\bibitem[{{Bonamente} {et~al.}(2022){Bonamente}, {Mirakhor}, {Lieu}, \& {Walker}}]{bonamente2022c}
{Bonamente}, M., {Mirakhor}, M., {Lieu}, R., \& {Walker}, S. 2022, \mnras, 514, 416

\bibitem[{{Bonamente} \& {Nevalainen}(2011)}]{bonamente2011}
{Bonamente}, M. \& {Nevalainen}, J. 2011, \apj, 738, 149

\bibitem[{{Bonnaire} {et~al.}(2020){Bonnaire}, {Aghanim}, {Decelle}, \& {Douspis}}]{Bonnaire2020}
{Bonnaire}, T., {Aghanim}, N., {Decelle}, A., \& {Douspis}, M. 2020, \aap, 637, A18

\bibitem[{{Bowyer} {et~al.}(1999){Bowyer}, {Bergh{\"o}fer}, \& {Korpela}}]{bowyer1999}
{Bowyer}, S., {Bergh{\"o}fer}, T.~W., \& {Korpela}, E.~J. 1999, \apj, 526, 592

\bibitem[{{Bowyer} {et~al.}(1996){Bowyer}, {Lampton}, \& {Lieu}}]{bowyer1996}
{Bowyer}, S., {Lampton}, M., \& {Lieu}, R. 1996, Science, 274, 1338

\bibitem[{{Braspenning} {et~al.}(2024){Braspenning}, {Schaye}, {Schaller}, {McCarthy}, {Kay}, {Helly}, {Kugel}, {Elbers}, {Frenk}, {Kwan}, {Salcido}, {van Daalen}, \& {Vandenbroucke}}]{Braspenning2023}
{Braspenning}, J., {Schaye}, J., {Schaller}, M., {et~al.} 2024, \mnras, 533, 2656

\bibitem[{{Cen}(1998)}]{cen1998}
{Cen}, R. 1998, \apjl, 498, L99

\bibitem[{{Cheng} {et~al.}(2005){Cheng}, {Borgani}, {Tozzi}, {Tornatore}, {Diaferio}, {Dolag}, {He}, {Moscardini}, {Murante}, \& {Tormen}}]{cheng2005}
{Cheng}, L.-M., {Borgani}, S., {Tozzi}, P., {et~al.} 2005, \aap, 431, 405

\bibitem[{{Churazov} {et~al.}(2023{\natexlab{a}}){Churazov}, {Khabibullin}, {Bykov}, {Lyskova}, \& {Sunyaev}}]{churazov2023b}
{Churazov}, E., {Khabibullin}, I., {Bykov}, A.~M., {Lyskova}, N., \& {Sunyaev}, R. 2023{\natexlab{a}}, \aap, 670, A156

\bibitem[{{Churazov} {et~al.}(2023{\natexlab{b}}){Churazov}, {Khabibullin}, {Dolag}, {Lyskova}, \& {Sunyaev}}]{churazov2023}
{Churazov}, E., {Khabibullin}, I.~I., {Dolag}, K., {Lyskova}, N., \& {Sunyaev}, R.~A. 2023{\natexlab{b}}, \mnras, 523, 1209

\bibitem[{{Dav{\'e}} {et~al.}(2001){Dav{\'e}}, {Cen}, {Ostriker}, {Bryan}, {Hernquist}, {Katz}, {Weinberg}, {Norman}, \& {O'Shea}}]{dave2001}
{Dav{\'e}}, R., {Cen}, R., {Ostriker}, J.~P., {et~al.} 2001, \apj, 552, 473

\bibitem[{{Dickey} \& {Lockman}(1990)}]{dickey1990}
{Dickey}, J.~M. \& {Lockman}, F.~J. 1990, \araa, 28, 215

\bibitem[{{Dietl} {et~al.}(2024){Dietl}, {Pacaud}, {Reiprich}, {Veronica}, {Migkas}, {Spinelli}, {Dolag}, \& {Seidel}}]{Dietl2024}
{Dietl}, J., {Pacaud}, F., {Reiprich}, T.~H., {et~al.} 2024, \aap, 691, A286

\bibitem[{{Dolag} {et~al.}(2016){Dolag}, {Komatsu}, \& {Sunyaev}}]{Magneticum}
{Dolag}, K., {Komatsu}, E., \& {Sunyaev}, R. 2016, \mnras, 463, 1797

\bibitem[{{Eckert} {et~al.}(2015){Eckert}, {Jauzac}, {Shan}, {Kneib}, {Erben}, {Israel}, {Jullo}, {Klein}, {Massey}, {Richard}, \& {Tchernin}}]{eckert2015}
{Eckert}, D., {Jauzac}, M., {Shan}, H., {et~al.} 2015, \nat, 528, 105

\bibitem[{{Finoguenov} {et~al.}(2003){Finoguenov}, {Briel}, \& {Henry}}]{finoguenov2003}
{Finoguenov}, A., {Briel}, U.~G., \& {Henry}, J.~P. 2003, \aap, 410, 777

\bibitem[{{Gal{\'a}rraga-Espinosa} {et~al.}(2020){Gal{\'a}rraga-Espinosa}, {Aghanim}, {Langer}, {Gouin}, \& {Malavasi}}]{galarraga2020}
{Gal{\'a}rraga-Espinosa}, D., {Aghanim}, N., {Langer}, M., {Gouin}, C., \& {Malavasi}, N. 2020, \aap, 641, A173

\bibitem[{{Gal{\'a}rraga-Espinosa} {et~al.}(2021){Gal{\'a}rraga-Espinosa}, {Aghanim}, {Langer}, \& {Tanimura}}]{galarraga2021}
{Gal{\'a}rraga-Espinosa}, D., {Aghanim}, N., {Langer}, M., \& {Tanimura}, H. 2021, \aap, 649, A117

\bibitem[{{Gallo} {et~al.}(2024){Gallo}, {Aghanim}, {Gouin}, {Eckert}, {Douspis}, {Paste}, \& {Bonnaire}}]{Gallo24}
{Gallo}, S., {Aghanim}, N., {Gouin}, C., {et~al.} 2024, \aap, 692, A200

\bibitem[{{Ghirardini} {et~al.}(2018){Ghirardini}, {Ettori}, {Eckert}, {Molendi}, {Gastaldello}, {Pointecouteau}, {Hurier}, \& {Bourdin}}]{Ghirardini2018}
{Ghirardini}, V., {Ettori}, S., {Eckert}, D., {et~al.} 2018, \aap, 614, A7

\bibitem[{{Gouin} {et~al.}(2023){Gouin}, {Bonamente}, {Gal{\'a}rraga-Espinosa}, {Walker}, \& {Mirakhor}}]{gouin2023}
{Gouin}, C., {Bonamente}, M., {Gal{\'a}rraga-Espinosa}, D., {Walker}, S., \& {Mirakhor}, M. 2023, \aap, 680, A94

\bibitem[{{Gouin} {et~al.}(2021){Gouin}, {Bonnaire}, \& {Aghanim}}]{gouin2021}
{Gouin}, C., {Bonnaire}, T., \& {Aghanim}, N. 2021, \aap, 651, A56

\bibitem[{{Gouin} {et~al.}(2022){Gouin}, {Gallo}, \& {Aghanim}}]{gouin2022}
{Gouin}, C., {Gallo}, S., \& {Aghanim}, N. 2022, \aap, 664, A198

\bibitem[{{Hahn} {et~al.}(2015){Hahn}, {Angulo}, \& {Abel}}]{Hahn2015}
{Hahn}, O., {Angulo}, R.~E., \& {Abel}, T. 2015, \mnras, 454, 3920

\bibitem[{{Hahn} {et~al.}(2017){Hahn}, {Martizzi}, {Wu}, {Evrard}, {Teyssier}, \& {Wechsler}}]{Hahn2017}
{Hahn}, O., {Martizzi}, D., {Wu}, H.-Y., {et~al.} 2017, \mnras, 470, 166

\bibitem[{{HI4PI Collaboration} {et~al.}(2016){HI4PI Collaboration}, {Ben Bekhti}, {Fl{\"o}er}, {Keller}, {Kerp}, {Lenz}, {Winkel}, {Bailin}, {Calabretta}, {Dedes}, {Ford}, {Gibson}, {Haud}, {Janowiecki}, {Kalberla}, {Lockman}, {McClure-Griffiths}, {Murphy}, {Nakanishi}, {Pisano}, \& {Staveley-Smith}}]{HI4PI2016}
{HI4PI Collaboration}, {Ben Bekhti}, N., {Fl{\"o}er}, L., {et~al.} 2016, \aap, 594, A116

\bibitem[{{Kaastra} {et~al.}(2003){Kaastra}, {Lieu}, {Tamura}, {Paerels}, \& {den Herder}}]{kaastra2003}
{Kaastra}, J.~S., {Lieu}, R., {Tamura}, T., {Paerels}, F.~B.~S., \& {den Herder}, J.~W. 2003, \aap, 397, 445

\bibitem[{{Kalberla} {et~al.}(2005){Kalberla}, {Burton}, {Hartmann}, {Arnal}, {Bajaja}, {Morras}, \& {P{\"o}ppel}}]{kalberla2005}
{Kalberla}, P.~M.~W., {Burton}, W.~B., {Hartmann}, D., {et~al.} 2005, \aap, 440, 775

\bibitem[{{Kotecha} {et~al.}(2022){Kotecha}, {Welker}, {Zhou}, {Wadsley}, {Kraljic}, {Sorce}, {Rasia}, {Roberts}, {Gray}, {Yepes}, \& {Cui}}]{Kotecha2022}
{Kotecha}, S., {Welker}, C., {Zhou}, Z., {et~al.} 2022, \mnras, 512, 926

\bibitem[{{Kull} \& {B{\"o}hringer}(1999)}]{kull1999}
{Kull}, A. \& {B{\"o}hringer}, H. 1999, \aap, 341, 23

\bibitem[{{Landry} {et~al.}(2013){Landry}, {Bonamente}, {Giles}, {Maughan}, {Joy}, \& {Murray}}]{landry2013}
{Landry}, D., {Bonamente}, M., {Giles}, P., {et~al.} 2013, \mnras, 433, 2790

\bibitem[{{Lebeau} {et~al.}(2024){Lebeau}, {Ettori}, {Aghanim}, \& {Sorce}}]{Lebeau2024}
{Lebeau}, T., {Ettori}, S., {Aghanim}, N., \& {Sorce}, J.~G. 2024, \aap, 689, A19

\bibitem[{{Lebeau} {et~al.}(2025){Lebeau}, {Zaroubi}, {Aghanim}, {Sorce}, \& {Langer}}]{Lebeau2025}
{Lebeau}, T., {Zaroubi}, S., {Aghanim}, N., {Sorce}, J.~G., \& {Langer}, M. 2025, \aap, 704, A14

\bibitem[{{Leccardi} \& {Molendi}(2008)}]{leccardi2008}
{Leccardi}, A. \& {Molendi}, S. 2008, \aap, 486, 359

\bibitem[{{Li} {et~al.}(2025){Li}, {Cui}, {Liu}, {Wang}, {Srivastava}, {Dave}, \& {Pearce}}]{Li2025}
{Li}, R., {Cui}, W., {Liu}, A., {et~al.} 2025, \aap, 701, A37

\bibitem[{{Lieu} {et~al.}(1996{\natexlab{a}}){Lieu}, {Mittaz}, {Bowyer}, {Breen}, {Lockman}, {Murphy}, \& {Hwang}}]{lieu1996b}
{Lieu}, R., {Mittaz}, J.~P.~D., {Bowyer}, S., {et~al.} 1996{\natexlab{a}}, Science, 274, 1335

\bibitem[{{Lieu} {et~al.}(1996{\natexlab{b}}){Lieu}, {Mittaz}, {Bowyer}, {Lockman}, {Hwang}, \& {Schmitt}}]{lieu1996a}
{Lieu}, R., {Mittaz}, J.~P.~D., {Bowyer}, S., {et~al.} 1996{\natexlab{b}}, \apjl, 458, L5

\bibitem[{{Liu} {et~al.}(2022){Liu}, {Bulbul}, {Ghirardini}, {Liu}, {Klein}, {Clerc}, {{\"O}zsoy}, {Ramos-Ceja}, {Pacaud}, {Comparat}, {Okabe}, {Bahar}, {Biffi}, {Brunner}, {Br{\"u}ggen}, {Buchner}, {Ider Chitham}, {Chiu}, {Dolag}, {Gatuzz}, {Gonzalez}, {Hoang}, {Lamer}, {Merloni}, {Nandra}, {Oguri}, {Ota}, {Predehl}, {Reiprich}, {Salvato}, {Schrabback}, {Sanders}, {Seppi}, \& {Thibaud}}]{liu2022}
{Liu}, A., {Bulbul}, E., {Ghirardini}, V., {et~al.} 2022, \aap, 661, A2

\bibitem[{{Lyskova} {et~al.}(2023){Lyskova}, {Churazov}, {Khabibullin}, {Burenin}, {Starobinsky}, \& {Sunyaev}}]{lyskova2023}
{Lyskova}, N., {Churazov}, E., {Khabibullin}, I.~I., {et~al.} 2023, \mnras, 525, 898

\bibitem[{{Mantz} {et~al.}(2022){Mantz}, {Morris}, {Allen}, {Canning}, {Baumont}, {Benson}, {Bleem}, {Ehlert}, {Floyd}, {Herbonnet}, {Kelly}, {Liang}, {von der Linden}, {McDonald}, {Rapetti}, {Schmidt}, {Werner}, \& {Wright}}]{mantz2022}
{Mantz}, A.~B., {Morris}, R.~G., {Allen}, S.~W., {et~al.} 2022, \mnras, 510, 131

\bibitem[{{Marini} {et~al.}(2025){Marini}, {Popesso}, {Dolag}, {Biffi}, {Vladutescu-Zopp}, {Castro}, {Toptun}, {de Is{\'\i}dio}, {Dev}, {Mazengo}, {Comparat}, {Gouin}, {Malavasi}, {Merloni}, {Mroczkowski}, {Ponti}, {Shreeram}, \& {Zhang}}]{Marini2025}
{Marini}, I., {Popesso}, P., {Dolag}, K., {et~al.} 2025, \aap, 698, A191

\bibitem[{{Marini} {et~al.}(2024){Marini}, {Popesso}, {Lamer}, {Dolag}, {Biffi}, {Vladutescu-Zopp}, {Dev}, {Toptun}, {Bulbul}, {Comparat}, {Malavasi}, {Merloni}, {Mroczkowski}, {Ponti}, {Seppi}, {Shreeram}, \& {Zhang}}]{Marini2024}
{Marini}, I., {Popesso}, P., {Lamer}, G., {et~al.} 2024, \aap, 689, A7

\bibitem[{{Martizzi} {et~al.}(2019){Martizzi}, {Vogelsberger}, {Artale}, {Haider}, {Torrey}, {Marinacci}, {Nelson}, {Pillepich}, {Weinberger}, {Hernquist}, {Naiman}, \& {Springel}}]{martizzi2019}
{Martizzi}, D., {Vogelsberger}, M., {Artale}, M.~C., {et~al.} 2019, \mnras, 486, 3766

\bibitem[{{Mernier} {et~al.}(2017){Mernier}, {de Plaa}, {Kaastra}, {Zhang}, {Akamatsu}, {Gu}, {Kosec}, {Mao}, {Pinto}, {Reiprich}, {Sanders}, {Simionescu}, \& {Werner}}]{mernier2017}
{Mernier}, F., {de Plaa}, J., {Kaastra}, J.~S., {et~al.} 2017, \aap, 603, A80

\bibitem[{{Mirakhor} \& {Walker}(2020)}]{mirakhor2020}
{Mirakhor}, M.~S. \& {Walker}, S.~A. 2020, \mnras, 497, 3204

\bibitem[{{Mirakhor} {et~al.}(2022){Mirakhor}, {Walker}, \& {Runge}}]{mirakhor2022}
{Mirakhor}, M.~S., {Walker}, S.~A., \& {Runge}, J. 2022, \mnras, 509, 1109

\bibitem[{{Mohapatra} {et~al.}(2022){Mohapatra}, {Jetti}, {Sharma}, \& {Federrath}}]{Mohapatra2022}
{Mohapatra}, R., {Jetti}, M., {Sharma}, P., \& {Federrath}, C. 2022, \mnras, 510, 2327

\bibitem[{{Nagai} \& {Lau}(2011)}]{nagai2011}
{Nagai}, D. \& {Lau}, E.~T. 2011, \apjl, 731, L10

\bibitem[{{Nelson} {et~al.}(2024){Nelson}, {Pillepich}, {Ayromlou}, {Lee}, {Lehle}, {Rohr}, \& {Truong}}]{Nelson2023}
{Nelson}, D., {Pillepich}, A., {Ayromlou}, M., {et~al.} 2024, \aap, 686, A157

\bibitem[{{Nelson} {et~al.}(2019){Nelson}, {Springel}, {Pillepich}, {Rodriguez-Gomez}, {Torrey}, {Genel}, {Vogelsberger}, {Pakmor}, {Marinacci}, {Weinberger}, {Kelley}, {Lovell}, {Diemer}, \& {Hernquist}}]{TNG2}
{Nelson}, D., {Springel}, V., {Pillepich}, A., {et~al.} 2019, Computational Astrophysics and Cosmology, 6, 2

\bibitem[{{Nevalainen} {et~al.}(2007){Nevalainen}, {Bonamente}, \& {Kaastra}}]{nevalainen2007}
{Nevalainen}, J., {Bonamente}, M., \& {Kaastra}, J. 2007, \apj, 656, 733

\bibitem[{{Nevalainen} {et~al.}(2003){Nevalainen}, {Lieu}, {Bonamente}, \& {Lumb}}]{nevalainen2003}
{Nevalainen}, J., {Lieu}, R., {Bonamente}, M., \& {Lumb}, D. 2003, \apj, 584, 716

\bibitem[{{Pillepich} {et~al.}(2018){Pillepich}, {Springel}, {Nelson}, {Genel}, {Naiman}, {Pakmor}, {Hernquist}, {Torrey}, {Vogelsberger}, {Weinberger}, \& {Marinacci}}]{TNG}
{Pillepich}, A., {Springel}, V., {Nelson}, D., {et~al.} 2018, \mnras, 473, 4077

\bibitem[{{Planelles} {et~al.}(2017){Planelles}, {Fabjan}, {Borgani}, {Murante}, {Rasia}, {Biffi}, {Truong}, {Ragone-Figueroa}, {Granato}, {Dolag}, {Pierpaoli}, {Beck}, {Steinborn}, \& {Gaspari}}]{Planelles2017}
{Planelles}, S., {Fabjan}, D., {Borgani}, S., {et~al.} 2017, \mnras, 467, 3827

\bibitem[{{Plucinsky} {et~al.}(1993){Plucinsky}, {Snowden}, {Briel}, {Hasinger}, \& {Pfeffermann}}]{plucinsky1993}
{Plucinsky}, P.~P., {Snowden}, S.~L., {Briel}, U.~G., {Hasinger}, G., \& {Pfeffermann}, E. 1993, \apj, 418, 519

\bibitem[{{Power} {et~al.}(2020){Power}, {Elahi}, {Welker}, {Knebe}, {Pearce}, {Yepes}, {Dav{\'e}}, {Kay}, {McCarthy}, {Puchwein}, {Borgani}, {Cunnama}, {Cui}, \& {Schaye}}]{Power2020}
{Power}, C., {Elahi}, P.~J., {Welker}, C., {et~al.} 2020, \mnras, 491, 3923

\bibitem[{{Rasia} {et~al.}(2014){Rasia}, {Lau}, {Borgani}, {Nagai}, {Dolag}, {Avestruz}, {Granato}, {Mazzotta}, {Murante}, {Nelson}, \& {Ragone-Figueroa}}]{rasia2014}
{Rasia}, E., {Lau}, E.~T., {Borgani}, S., {et~al.} 2014, \apj, 791, 96

\bibitem[{{Reiprich} {et~al.}(2021){Reiprich}, {Veronica}, {Pacaud}, {Ramos-Ceja}, {Ota}, {Sanders}, {Kara}, {Erben}, {Klein}, {Erler}, {Kerp}, {Hoang}, {Br{\"u}ggen}, {Marvil}, {Rudnick}, {Biffi}, {Dolag}, {Aschersleben}, {Basu}, {Brunner}, {Bulbul}, {Dennerl}, {Eckert}, {Freyberg}, {Gatuzz}, {Ghirardini}, {K{\"a}fer}, {Merloni}, {Migkas}, {Nandra}, {Predehl}, {Robrade}, {Salvato}, {Whelan}, {Diaz-Ocampo}, {Hernandez-Lang}, {Zenteno}, {Brown}, {Collier}, {Diego}, {Hopkins}, {Kapinska}, {Koribalski}, {Mroczkowski}, {Norris}, {O'Brien}, \& {Vardoulaki}}]{reiprich2021}
{Reiprich}, T.~H., {Veronica}, A., {Pacaud}, F., {et~al.} 2021, \aap, 647, A2

\bibitem[{{Rost} {et~al.}(2021){Rost}, {Kuchner}, {Welker}, {Pearce}, {Stasyszyn}, {Gray}, {Cui}, {Dave}, {Knebe}, {Yepes}, \& {Rasia}}]{Rost2021}
{Rost}, A., {Kuchner}, U., {Welker}, C., {et~al.} 2021, \mnras, 502, 714

\bibitem[{{Rost} {et~al.}(2024){Rost}, {Nuza}, {Stasyszyn}, {Kuchner}, {Hoeft}, {Welker}, {Pearce}, {Gray}, {Knebe}, {Cui}, \& {Yepes}}]{Rost2024}
{Rost}, A.~M., {Nuza}, S.~E., {Stasyszyn}, F., {et~al.} 2024, \mnras, 527, 1301

\bibitem[{{Rybicki} \& {Lightman}(1979)}]{rybicki1979}
{Rybicki}, G.~B. \& {Lightman}, A.~P. 1979, Radiative Processes in Astrophysics (New York, Wiley-Interscience)

\bibitem[{{Sakelliou} \& {Ponman}(2004)}]{sakelliou2004}
{Sakelliou}, I. \& {Ponman}, T.~J. 2004, \mnras, 351, 1439

\bibitem[{{Shi} {et~al.}(2020){Shi}, {Nagai}, {Aung}, \& {Wetzel}}]{Shi2020}
{Shi}, X., {Nagai}, D., {Aung}, H., \& {Wetzel}, A. 2020, \mnras, 495, 784

\bibitem[{{Shreeram} {et~al.}(2025{\natexlab{a}}){Shreeram}, {Comparat}, {Merloni}, {Ponti}, {Popesso}, {Zhang}, {Nandra}, {Salvato}, {Marini}, {Buchner}, {Locatelli}, \& {Igo}}]{Soumya_model}
{Shreeram}, S., {Comparat}, J., {Merloni}, A., {et~al.} 2025{\natexlab{a}}, \aap, 703, A137

\bibitem[{{Shreeram} {et~al.}(2025{\natexlab{b}}){Shreeram}, {Comparat}, {Merloni}, {Zhang}, {Ponti}, {Nandra}, {ZuHone}, {Marini}, {Vladutescu-Zopp}, {Popesso}, {Pakmor}, {Seppi}, {Peroux}, \& {Sorini}}]{Soumya_LC}
{Shreeram}, S., {Comparat}, J., {Merloni}, A., {et~al.} 2025{\natexlab{b}}, \aap, 697, A22

\bibitem[{{Snowden} {et~al.}(1997){Snowden}, {Egger}, {Freyberg}, {McCammon}, {Plucinsky}, {Sanders}, {Schmitt}, {Truemper}, \& {Voges}}]{snowden1997}
{Snowden}, S.~L., {Egger}, R., {Freyberg}, M.~J., {et~al.} 1997, \apj, 485, 125

\bibitem[{{Snowden} {et~al.}(1995){Snowden}, {Freyberg}, {Plucinsky}, {Schmitt}, {Truemper}, {Voges}, {Edgar}, {McCammon}, \& {Sanders}}]{snowden1995}
{Snowden}, S.~L., {Freyberg}, M.~J., {Plucinsky}, P.~P., {et~al.} 1995, \apj, 454, 643

\bibitem[{{Snowden} {et~al.}(1994){Snowden}, {McCammon}, {Burrows}, \& {Mendenhall}}]{snowden1994}
{Snowden}, S.~L., {McCammon}, D., {Burrows}, D.~N., \& {Mendenhall}, J.~A. 1994, \apj, 424, 714

\bibitem[{{Tanimura} {et~al.}(2022){Tanimura}, {Aghanim}, {Douspis}, \& {Malavasi}}]{Tanimura2022}
{Tanimura}, H., {Aghanim}, N., {Douspis}, M., \& {Malavasi}, N. 2022, \aap, 667, A161

\bibitem[{{Tuominen} {et~al.}(2023){Tuominen}, {Nevalainen}, {Hein{\"a}m{\"a}ki}, {Tempel}, {Wijers}, {Bonamente}, {Aragon-Calvo}, \& {Finoguenov}}]{Tuominen2023}
{Tuominen}, T., {Nevalainen}, J., {Hein{\"a}m{\"a}ki}, P., {et~al.} 2023, \aap, 671, A103

\bibitem[{{Tuominen} {et~al.}(2021){Tuominen}, {Nevalainen}, {Tempel}, {Kuutma}, {Wijers}, {Schaye}, {Hein{\"a}m{\"a}ki}, {Bonamente}, \& {Ganeshaiah Veena}}]{tuominen2021}
{Tuominen}, T., {Nevalainen}, J., {Tempel}, E., {et~al.} 2021, \aap, 646, A156

\bibitem[{{Vall{\'e}s-P{\'e}rez} {et~al.}(2021){Vall{\'e}s-P{\'e}rez}, {Planelles}, \& {Quilis}}]{VallesPerez2021}
{Vall{\'e}s-P{\'e}rez}, D., {Planelles}, S., \& {Quilis}, V. 2021, \mnras, 504, 510

\bibitem[{{Vogelsberger} {et~al.}(2018){Vogelsberger}, {Marinacci}, {Torrey}, {Genel}, {Springel}, {Weinberger}, {Pakmor}, {Hernquist}, {Naiman}, {Pillepich}, \& {Nelson}}]{Vogelsberger2018}
{Vogelsberger}, M., {Marinacci}, F., {Torrey}, P., {et~al.} 2018, \mnras, 474, 2073

\bibitem[{{Voges} {et~al.}(1999){Voges}, {Aschenbach}, {Boller}, {Br{\"a}uninger}, {Briel}, {Burkert}, {Dennerl}, {Englhauser}, {Gruber}, {Haberl}, {Hartner}, {Hasinger}, {K{\"u}rster}, {Pfeffermann}, {Pietsch}, {Predehl}, {Rosso}, {Schmitt}, {Tr{\"u}mper}, \& {Zimmermann}}]{voges1999}
{Voges}, W., {Aschenbach}, B., {Boller}, T., {et~al.} 1999, \aap, 349, 389

\bibitem[{{Vurm} {et~al.}(2023){Vurm}, {Nevalainen}, {Hong}, {Bah{\'e}}, {Dalla Vecchia}, \& {Hein{\"a}m{\"a}ki}}]{Vurm2023}
{Vurm}, I., {Nevalainen}, J., {Hong}, S.~E., {et~al.} 2023, \aap, 673, A62

\bibitem[{{Walker} {et~al.}(2012){Walker}, {Fabian}, {Sanders}, {George}, \& {Tawara}}]{walker2012a}
{Walker}, S.~A., {Fabian}, A.~C., {Sanders}, J.~S., {George}, M.~R., \& {Tawara}, Y. 2012, \mnras, 422, 3503

\bibitem[{{Wang} {et~al.}(2024){Wang}, {Wang}, {Guo}, {Kang}, {Libeskind}, {Gal{\'a}rraga-Espinosa}, {Springel}, {Kannan}, {Hernquist}, {Pakmor}, {Yu}, {Bose}, {Guo}, {Yu}, \& {Hern{\'a}ndez-Aguayo}}]{Wang2024}
{Wang}, W., {Wang}, P., {Guo}, H., {et~al.} 2024, \mnras, 532, 4604

\bibitem[{{Werner} {et~al.}(2008){Werner}, {Finoguenov}, {Kaastra}, {Simionescu}, {Dietrich}, {Vink}, \& {B{\"o}hringer}}]{werner2008}
{Werner}, N., {Finoguenov}, A., {Kaastra}, J.~S., {et~al.} 2008, \aap, 482, L29

\bibitem[{{Zappacosta} {et~al.}(2005){Zappacosta}, {Maiolino}, {Mannucci}, {Gilli}, \& {Schuecker}}]{zappacosta2005}
{Zappacosta}, L., {Maiolino}, R., {Mannucci}, F., {Gilli}, R., \& {Schuecker}, P. 2005, \mnras, 357, 929

\bibitem[{{Zhang} {et~al.}(2024{\natexlab{a}}){Zhang}, {Zhuravleva}, {Markevitch}, {ZuHone}, {Mernier}, {Biffi}, {Bogd{\'a}n}, {Chakraborty}, {Churazov}, {Dolag}, {Ettori}, {Forman}, {Hernquist}, {Jones}, {Khabibullin}, {Kilbourne}, {Kraft}, {Lau}, {Lin}, {Nagai}, {Nelson}, {Ogorza{\l}ek}, {Rasia}, {Sarkar}, {Simionescu}, {Su}, {Vogelsberger}, \& {Walker}}]{Zhang2024}
{Zhang}, C., {Zhuravleva}, I., {Markevitch}, M., {et~al.} 2024{\natexlab{a}}, \mnras, 530, 4234

\bibitem[{{Zhang} {et~al.}(2024{\natexlab{b}}){Zhang}, {Bulbul}, {Malavasi}, {Ghirardini}, {Comparat}, {Kluge}, {Liu}, {Merloni}, {Zhang}, {Bahar}, {Artis}, {Sanders}, {Garrel}, {Balzer}, {Br{\"u}ggen}, {Freyberg}, {Gatuzz}, {Grandis}, {Krippendorf}, {Nandra}, {Ponti}, {Ramos-Ceja}, {Predehl}, {Reiprich}, {Veronica}, {Yeung}, \& {Zelmer}}]{Zhang2024_erosita}
{Zhang}, X., {Bulbul}, E., {Malavasi}, N., {et~al.} 2024{\natexlab{b}}, \aap, 691, A234

\bibitem[{{Zhao} {et~al.}(2025){Zhao}, {Xu}, {Liu}, {Zhang}, {Ji}, {Chang}, {Hu}, {Werner}, {Zhang}, {Cui}, \& {Wu}}]{Zhao2025}
{Zhao}, Y., {Xu}, H., {Liu}, A., {et~al.} 2025, \aap, 695, A15

\end{thebibliography}
\bibliographystyle{aa}
\end{document}